\documentclass{ar-1col}

\usepackage[comma]{natbib}
\usepackage{amssymb,amsmath}
\usepackage{amsfonts,subfigure,graphics,graphicx}
\usepackage{multirow}
\usepackage{amsthm,bm}

\usepackage{geometry}
\begin{document}

\markboth{Roy}{convergence diagnostics for MCMC}

\title{Convergence diagnostics for Markov chain Monte Carlo}

\author{Vivekananda Roy 
\affil{ Department of statistics, Iowa state University\\
Ames, Iowa, USA; email: vroy@iastate.edu}
}

\begin{abstract}
  Markov chain Monte Carlo (MCMC) is one of the most useful approaches
  to scientific computing because of its flexible construction, ease
  of use and generality. Indeed, MCMC is indispensable for performing
  Bayesian analysis. Two critical
  questions that MCMC practitioners need to address are where to start
  and when to stop the simulation. Although a great amount of research
  has gone into establishing convergence criteria and stopping rules
  with sound theoretical foundation, in practice, MCMC users often
  decide convergence by applying empirical diagnostic tools. This
  review article discusses the most widely used MCMC convergence
  diagnostic tools. Some recently proposed stopping rules with
  firm theoretical footing are also presented. The convergence diagnostics
  and stopping rules are illustrated using three detailed examples.
\end{abstract}


\maketitle


\section{INTRODUCTION}
\label{sec:int}
Markov chain Monte Carlo (MCMC) methods are now routinely used to fit
complex models in diverse disciplines. A Google search for
``Markov chain Monte Carlo'' returns more than 11.5 million hits. The popularity of
MCMC is mainly due to its widespread usage in computational physics and
Bayesian statistics, although it is also used in frequentist inference
\cite[see e.g.][]{geye:thomp:1995, chri:2004}.

The fundamental idea of MCMC is that if simulating from a target
density $\pi$ is difficult so that the ordinary Monte Carlo method
based on independent and identically distributed (iid) samples cannot
be used for making inference on $\pi$, it may be possible to construct
a Markov chain $\{X_n\}_{n \ge 0}$ with stationary density $\pi$ for
forming Monte Carlo estimators. An introduction to construction of
such Markov chains, including the Gibbs sampler and the
Metropolis-Hasting (MH) sampler, is provided by \cite{geye:2011}
\cite[see also][]{robe:case:2004}. General purpose MH algorithms are
available in the R packages mcmc \citep{r:mcmc} and MCMCpack
\citep{R:mcmcp}. There are several R \citep{r} packages implementing
specific MCMC algorithms for a number of statistical models [see
e.g. MCMCpack \citep{R:mcmcp}, MCMCglmm \citep{r:MCMCglmm}, geoBayes
\citep{r:geoBayes}]. Here, we do not discuss development of MCMC
algorithms, but rather focus on analyzing the Markov chain obtained
from running such an algorithm for determining its convergence.

Two important issues that must be addressed while implementing MCMC
are where to start and when to stop the algorithm. As we discuss now,
these two tasks are related to determining convergence of the
underlying Markov chain to stationarity and convergence of Monte Carlo
estimators to population quantities, respectively. It is known that
under some standard conditions on the Markov chain, for any initial
value, the distribution of $X_n$ converges to the stationary
distribution as $n \rightarrow \infty$ (see
e.g. \citet[][chap. 13]{meyn:twee:1993}, \cite{robe:rose:2004}). Since
$X_0 \not\sim \pi$ and MCMC algorithms produce (serially) correlated
samples, the further the initial distribution from $\pi$, the longer
it takes for $X_n$ to approximate $\pi$. In particular, if the
initial value is not in a high-density ($\pi$) region, the
samples at the earlier iterations may not be close to the target
distribution. In such cases, a common practice is to discard early
realizations in the chain and start collecting samples only after the
effect of the initial value has (practically) worn off. The main
idea behind this method, known as {\it burn-in}, is to use samples
only after the Markov chain gets sufficiently close to the stationary
distribution, although its usefulness for Monte Carlo estimation has
been questioned in the MCMC community \citep{geye:2011}. Thus,
ideally, MCMC algorithms should be initialized at a high-density region,
but if finding such areas is difficult, collection of Monte Carlo
samples can be started only after a certain iteration $n'$ when
approximately $X_{n'} \sim \pi$.

Once the starting value is determined, one needs to decide when to
stop the simulation. (Note that the starting value here refers to the
beginning of collection of samples as opposed to the initial value of
$X_0$ of the Markov chain, although these two values can be the same.)
Often the quantities of interest regarding the target density $\pi$
can be expressed as means of certain functions, say
$E_\pi g \equiv \int_{\mathcal{X}} g(x) \pi(x) dx$ where $g$ is a real
valued function. For example, appropriate choices of $g$ make
$E_\pi g$ different measures of location, spread, and other summary
features of $\pi$. Here, the support of the target density $\pi$ is
denoted by $\mathcal{X}$, which is generally $\mathbb{R}^d$ for some
$d \ge 1$, although it can be non-Euclidean as well.  We later in
Section~\ref{sec:diag} consider vector valued functions $g$ as
well. The MCMC estimator of the population mean $E_\pi g$ is the
sample average
$\bar{g}_{n', n} \equiv \sum_{i=n'+1}^{n} g(X_i)/(n-n')$. If no
burn-in is used then $n'=0$. It is known that usually
$\bar{g}_{n', n} \rightarrow E_\pi g $ as $n \rightarrow \infty$ (see
Section~\ref{sec:diag} for details). In practice, however, MCMC users
run the Markov chain for a finite $n^*$ number of iterations, thus
MCMC simulation should be stopped only when $\bar{g}_{n', n^*}$ has
sufficiently converged to $E_\pi g$. The accuracy of the time average
estimator $\bar{g}_{n', n}$ obviously depends on the quality of the
samples. Thus, when implementing MCMC methods, it is necessary to
wisely conclude Markov chain convergence, and subsequently determine
when to stop the simulation. In particular, while premature
termination of the simulation will most likely lead to inaccurate
inference, unnecessarily running longer chains is not desirable either
as it eats up resources.

By performing theoretical analysis on the underlying Markov chain, an
analytical upper bound on its distance to stationarity may be obtained
\citep{rose:1995}, which in turn can provide a rigorous method for
deciding MCMC convergence and thus finding $n'$
\citep{jone:hobe:2001}. Similarly, using a sample size calculation
based on an asymptotic distribution of the (appropriately scaled)
Monte Carlo error $\bar{g}_{n', n^*} - E_\pi g$, an {\it honest}
stopping value $n^*$ can be found. In the absence of such theoretical
analysis, often empirical diagnostic tools are used to check
convergence of MCMC samplers and estimators, although, as shown
through examples in Section~\ref{sec:exam}, these tools cannot
determine convergence with certainty.  Since early 1990s with the
increasing use of MCMC, a great deal of research effort has gone into
developing convergence diagnostic tools. These diagnostic methods can
be classified into several categories. For example, corresponding to
the two types of convergence mentioned before, some of these
diagnostic tools are designed to assess convergence of the Markov
chain to the stationary distribution, whereas others check for
convergence of the summary statistics like sample means and sample
quantiles to the corresponding population quantities. The available
MCMC diagnostic methods can be categorized according to other criteria
as well, for example, their level of theoretical foundation, if they
are suitable for checking joint convergence of multiple variables,
whether they are based on multiple (parallel) chains or a single chain
or both, if they are complemented by a visualization tool or not, if
they are based on moments and quantiles or the kernel density of the
observed chain, and so on. Several review articles on MCMC convergence
diagnostics are available in the literature \cite[see
e.g.][]{cowl:carl:1996, broo:robe:1998,
  meng:robe:guih:1999}. \cite{cowl:carl:1996} provide a description of
13 convergence diagnostics and summarize these according to the
different criteria mentioned above. While some of these methods are
widely used in practice, several new approaches have been proposed
since then. In this article, we review some of these tools that are
commonly used by MCMC practitioners or that we find promising.

\section{MCMC diagnostics}
\label{sec:diag}
As mentioned in the introduction, MCMC diagnostic tools are needed for
deciding convergence of Markov chains to the stationarity.  Also,
although in general the longer the chain is run the better Monte Carlo
estimates it produces, in practice, it is desirable to use some
stopping rules for prudent use of resources. In this section, we
describe some MCMC diagnostics that may be used for deciding Markov
chain convergence or stopping MCMC sampling. In the context of each
method, we also report if it is designed particularly for one of these
two objectives.

\subsection{Honest MCMC}
\label{sec:hone}
In this section, we describe some rigorous methods for finding $n'$
and $n^*$ mentioned in the introduction. Let $f_n$ be the density of
$X_n$. It is known that under some standard conditions \citep[see
e.g.][chap. 13]{meyn:twee:1993},
$\frac{1}{2} \int_{\mathcal{X}} |f_n(x) - \pi (x)| dx \downarrow 0$ as
$n \rightarrow \infty$, that is, $X_n$ converges in the total
variation (TV) norm to a random variable following
$\pi$. \cite{jone:hobe:2001} mention that a rigorous way of deciding
the convergence of the Markov chain to $\pi$ is by finding an iteration
number $n'$ such that
\begin{equation}
  \label{eq:totvar}
  \frac{1}{2} \int_{\mathcal{X}} |f_{n'}(x) - \pi (x)| dx < 0.01.
\end{equation}
(The cutoff value 0.01 is arbitrary and any predetermined precision
level can be used.)   \cite{jone:hobe:2001} propose to use the
smallest $n'$ for which (\ref{eq:totvar}) holds as the honest value
for burn-in.

The above-mentioned burn-in hinges on the TV norm in (\ref{eq:totvar})
which is generally not available. Constructing a quantitative bound to
the TV norm is also often difficult, although significant progress has
been made in this direction \citep{rose:1995, rose:2002, baxe:2005,
  andr:fort:viho:2015}.  In particular, a key tool for constructing a
quantitative bound to the TV norm is using the {\it drift} and {\it
  minorization} (d\&m) technique \citep{rose:1995}. The d\&m technique
has been successfully used to analyze a variety of MCMC algorithms
\cite[see e.g.][]{ fort:moul:robe:rose:2003, jone:hobe:2004,
  roy:hobe:2010, vats:2017}. The d\&m conditions, as we explain later
in this section, are also crucial to provide an honest way to check
convergence of MCMC estimators of popular summary measures like
moments and quantiles of the target distributions. Although we
consider the TV norm here, over the last few years, other metrics like
the Wasserstein distance have also been used to study Markov chain
convergence \cite[see e.g.][]{durm:moul:2015,
  qin:hobe:2019}. On the other hand, using Stein's method,
\cite{gorh:mack:2015} propose a computable discrepancy measure that
seems promising as it depends on the target only through the
derivative of $\log \pi$, and hence is appropriate in Bayesian
settings where the target is generally known up to the intractable
normalizing constant.

As in the Introduction, let a particular feature of the target density
be expressed as $E_\pi g$ where $g$ is a real valued function.
By the strong law of large numbers for Markov chains, it is known that
if $\{X_n\}_{n \ge 0}$ is appropriately {\it irreducible}, then
$\bar{g}_{n',n} \equiv \sum_{i=n'+1}^{n} g(X_i)/(n-n')$ is a strongly
consistent estimator of $E_\pi g$, that is,
$\bar{g}_{n',n} \rightarrow E_\pi g$ almost surely as
$n \rightarrow \infty$ for any fixed $n'$
\citep{asmu:glyn:2011}. Without loss of generality, we let $n' =0$
when discussing stopping rules, and for the ease of notation, we
simply write $\bar{g}_{n}$ for $\bar{g}_{0,n}$. The law of large numbers
justifies estimating $E_\pi g$ by the sample (time) average estimator
$\bar{g}_n$, as in the ordinary Monte Carlo. If a central limit
theorem (CLT) is available for $\bar{g}_n$ (that is, for the error
$\bar{g}_n - E_\pi g$) then a `sample size calculation' based on the
width of an interval estimator for $E_\pi g$ can be performed for
choosing an appropriate value for $n^*$. Indeed, under some regularity
conditions,
\begin{equation}
  \label{eq:clt}
  \sqrt{n} (\bar{g}_n - E_\pi g) \stackrel{d}{\rightarrow} N(0,
  \sigma^2_g) \;\;\mbox{as}\; n \rightarrow \infty,
\end{equation}
where
$\sigma^2_g \equiv \mbox{Var}_\pi (g(X_0)) + 2 \sum_{i=1}^{\infty}
\mbox{Cov}_\pi (g(X_0), g(X_i)) < \infty$; the subscript $\pi$
indicates that the expectations are calculated assuming
$X_0 \sim \pi$. (Note that, due to the autocorrelations present in a
Markov chain, $\sigma^2_g \neq \mbox{Var}_\pi (g(X_0)) = \lambda^2_g$,
say.) If $\widehat{\sigma}_{g, n}$ is a consistent estimator of
$\sigma_g$, then an estimator of the standard error of $\bar{g}_n$,
based on the sample size $n$ is $\widehat{\sigma}_{g, n}/\sqrt{n}$. Since the
standard error $\widehat{\sigma}_{g, n}/\sqrt{n}$ allows one to judge the
reliability of the MCMC estimate, it should always be reported along
with the point estimate $\bar{g}_n$. The standard error also leads to
a $100(1-\alpha)\%$ confidence interval for $E_\pi g$, namely
$\bar{g}_n \mp z_{\alpha/2} \widehat{\sigma}_{g, n}/\sqrt{n}$. Here
$z_{\alpha/2}$ is the $(1 -\alpha/2)$ quantile of the standard
normal distribution. The MCMC simulation can be stopped if the
half-width of the $100(1-\alpha)\%$ confidence interval
falls below a prespecified threshold, say $\epsilon$. \cite{jone:hobe:2001} refer to
this method as the honest way to stop the chain. Indeed, the fixed-width stopping rule (FWSR)
\citep{fleg:hara:jone:2008, jone:hara:caff:neat:2006} terminates
the simulation the first time after some user-specified $\tilde{n}$ iterations that
\begin{equation}
  \label{eq:fwsr}
t_{*} \frac{\widehat{\sigma}_{g, n}}{\sqrt{n}} + \frac{1}{n} \le \epsilon.
\end{equation}
Here, $t_{*}$ is an appropriate quantile.  The role of $\tilde{n}$ is
to make sure that the simulation is not stopped prematurely due to
poor estimate of $\widehat{\sigma}_{g, n}$. The value of $\tilde{n}$
should depend on the complexity of the problem. \cite{gong:fleg:2016}
suggest that using $\tilde{n} = 10^4$ works well in
practice. 

For validity of the honest stopping rule, a CLT (\ref{eq:clt})
for $\bar{g}_n$ needs to exist, and one would need a consistent
estimator $\widehat{\sigma}_{g, n}$ of $\sigma_g$. For the CLT to
hold, the TV norm in (\ref{eq:totvar}) needs to converge to zero at
certain rate \cite[see][for different conditions guaranteeing a Markov
chain CLT]{jone:2004}. The most common method of establishing a CLT
(\ref{eq:clt}) as well as providing a consistent
estimator of $\sigma_g$ has been by showing the Markov chain
$\{X_n\}_{n \ge 0}$ is {\it geometrically ergodic}, that is, the TV
norm (\ref{eq:totvar}) converges at an exponential rate
\citep{jone:hobe:2001, robe:rose:2004}. Generally, geometric
ergodicity of a Markov chain is proven by constructing an appropriate
d\&m condition \citep{rose:1995, roy:hobe:2010}. For estimation of
$\sigma^2_g$,  while \cite{mykl:tier:yu:1995}, and \cite{hobe:jone:pres:rose:2002}
 discuss regenerative submission method,
\cite{jone:hara:caff:neat:2006} and \cite{fleg:jone:2010} provide
consistent batch means and spectral variance methods. Availability of
a Markov chain CLT has been demonstrated for myriad MCMC algorithms
for common statistical models. Here we provide an incomplete
list: linear models \citep{roma:hobe:2012, roma:hobe:2015},
generalized linear models including the probit model
\citep{roy:hobe:2007, chak:khar:2017}, the popular logistic model
\citep{choi:hobe:2013, wang:roy:2018b} and the robit model
\citep{roy:2012b}, generalized linear mixed models including the
probit mixed model \citep{wang:roy:2018}, and the logistic mixed model
\citep{wang:roy:2018a}, quantile regression models
\citep{khar:hobe:2012}, multivariate regression models
\citep{roy:hobe:2010, hobe:jung:khar:qin:2018}, penalized regression
 and variable selection models \citep{khar:hobe:2013,
  roy:chak:2017, vats:2017}.

So far we have described the honest MCMC in the context of estimating
means of univariate functions. The method is applicable to estimation
of vector valued functions as well. In particular, if $g$ is a
$\mathbb{R}^p$ valued function, and if a CLT holds for $\bar{g}_n$,
that is, if
$ \sqrt{n} (\bar{g}_n - E_\pi g) \stackrel{d}{\rightarrow} N(0,
\Sigma_g) \;\;\mbox{as}\; n \rightarrow \infty,$ for some $p \times p$
covariance matrix $\Sigma_g$, then using a consistent estimator
$\widehat{\Sigma}_{g, n}$ of $\Sigma_g$, a $100(1-\alpha)\%$
asymptotic confidence region $C_\alpha(n)$ for $E_\pi g$ can be formed
\cite[for
details see][]{vats:fleg:jone:2019}. \cite{vats:fleg:jone:2019} propose
a fixed-volume stopping rule which terminates the simulation the first
time after $\tilde{n}$ iterations that
\[
  (\mbox{Vol}\{C_\alpha(n)\})^{1/p} + \frac{1}{n} \le \varepsilon,
  \]
where as in (\ref{eq:fwsr}), $\varepsilon$ is the user's desired level
of accuracy. Note that when
$p=1$, except the $1/n$ terms, the expression above is same as (\ref{eq:fwsr}) with
$\varepsilon = 2\epsilon$.  Honest MCMC can also be implemented for
estimation of the quantiles \citep{doss:fleg:jone:neat:2014}. In order
to reduce computational burden, the sequential stopping rules should
be checked only at every $l$ iterations where $l$ is appropriately
chosen. Finally, even if theoretical d\&m analysis is not carried out
establishing a Markov chain CLT, in practice, FWSR can be implemented
using the batch means and spectral variance estimators of
$\sigma_g (\Sigma_g) $ available in the R package mcmcse
\citep{R:mcmcse}.

\subsection{Relative fixed-width stopping rules}
\label{sec:relstop}

FWSR (described in Section~\ref{sec:hone}) explicitly address how well
the estimator $\bar{g}_n$ approximates $E_\pi g$.
\cite{fleg:gong:2015} and \cite{gong:fleg:2016} discuss relative
FWSR in the MCMC setting. \cite{fleg:gong:2015} consider a relative
magnitude rule that terminates the simulation when after $\tilde{n}$
iterations
$t_{*} \widehat{\sigma}_{g, n} n^{-1/2} + n^{-1} \le \epsilon
\bar{g}_n$. \cite{fleg:gong:2015} also consider a relative standard
deviation FWSR (SDFWSR) that terminates the simulation when after
$\tilde{n}$ iterations
$t_{*} \widehat{\sigma}_{g, n} n^{-1/2} + n^{-1} \le \epsilon
\widehat{\lambda}_{g, n}$, where $\widehat{\lambda}_{g, n}$ is a
strongly consistent estimator of the population standard deviation $\lambda_g$.
Asymptotic validity of the relative magnitude and relative standard deviation stopping
rules is established by \cite{glyn:whit:1992} and
\cite{fleg:gong:2015} respectively. This ensures that the simulation
will terminate in a finite time with probability $1$.

In Bayesian statistics applications, \cite{fleg:gong:2015} advocate
the use of relative SDFWSR. In the high-dimensional settings, that is,
where $g$ is a $\mathbb{R}^p$ valued function and $p$ is large,
without a priori knowledge of the magnitude of $E_\pi g$,
\cite{gong:fleg:2016} prefer relative SDFWSR over FWSR based on the marginal chains. In the
multivariate settings, \cite{vats:fleg:jone:2019} argue that stopping
rules based on $p$ marginal chains may not be appropriate as these
ignore cross-correlations between components and may be dictated by
the slowest mixing marginal chain. \cite{vats:fleg:jone:2019} propose
a multivariate relative standard deviation stopping rule involving
volume of the $100(1-\alpha)\%$ asymptotic confidence region, that is,
$\mbox{Vol}\{C_\alpha(n)\}$. Let $\widehat{\Lambda}_{g, n}$ be the sample
covariance matrix. \cite{vats:fleg:jone:2019} propose to stop the
simulation, the first time after $\tilde{n}$ iterations that
\begin{equation}
  \label{eq:mulrelsd}
    (\mbox{Vol}\{C_\alpha(n)\})^{1/p} + \frac{1}{n} \le \varepsilon (|\widehat{\Lambda}_{g, n}|)^{1/2p},
\end{equation}
where $| \cdot |$ denotes the determinant.
\subsection{Effective sample size}
\label{sec:ess}

For an MCMC-based estimator, effective sample size (ESS) is the number
of independent samples equivalent to (that is, having the same
standard error as) a set of correlated Markov chain samples. Although
ESS (based on $n$ correlated samples) is not uniquely defined, the
most common definition \citep{robe:case:2004} is
\[
  \mbox{ESS} = \frac{n}{1 + 2\sum_{i=1}^{\infty}\mbox{Corr}_\pi (g(X_0), g(X_i))}.
\]
\cite{gong:fleg:2016} rewrite the above definition as ESS
$= n \lambda_g^2/\sigma^2_g$. In the multivariate setting, that is,
when $g$ is $\mathbb{R}^p$ valued for some $p\ge 1$,
\cite{vats:fleg:jone:2019} define multivariate ESS (mESS) as
\begin{equation}
  \label{eq:mess}
  \mbox{mESS} = n\bigg(\frac{|\Lambda_g|}{|\Sigma_g|}\bigg)^{1/p},
\end{equation}
where $\Lambda_g$ is the population covariance matrix.  An approach to
terminate MCMC simulation is when
$\widehat{\mbox{ESS}} \; (\widehat{\mbox{mESS}}) $ takes a value
larger than a prespecified number, where
$\widehat{\mbox{ESS}} \;(\widehat{\mbox{mESS}})$ is a consistent
estimator of $\mbox{ESS}\; (\mbox{mESS})$. Indeed,
\cite{vats:fleg:jone:2019} mention that simulation can be terminated
the first time that
\begin{equation}
  \label{eq:messcut}
  \widehat{\mbox{mESS}} = n\bigg(\frac{|\widehat{\Lambda_{g, n|}}}{|\widehat{\Sigma_{g, n}}|}\bigg)^{1/p} \ge \frac{2^{2/p}\pi}{(p \Gamma(p/2))^{2/p}} \frac{\chi^2_{1-\alpha, p}}{\varepsilon^2},
  \end{equation}
  where $\varepsilon$ is the desired level of precision for the volume
  of the $100(1-\alpha)\%$ asymptotic confidence region, and
  $\chi^2_{1-\alpha, p}$ is the $(1- \alpha)$ quantile of
  $\chi^2_{p}$. This ESS stopping rule is (approximately) equivalent
  to the multivariate relative standard deviation stopping rule given
  in (\ref{eq:mulrelsd}) \cite[for details,
  see][]{vats:fleg:jone:2019}. Note that
  $\widehat{\mbox{ESS}} \; (\widehat{\mbox{mESS}})$ per unit time can
  be used to compare different MCMC algorithms (with the same
  stationary distribution) in terms of both computational and
  statistical efficiency. ESS is implemented in several R packages
  including coda \citep{R:coda} and mcmcse \citep{R:mcmcse}. In the
  mcmcse package, estimates of ESS both in univariate and multivariate
  settings are available. While \cite{gong:fleg:2016} and
  \cite{vats:fleg:jone:2019} provide a connection between ESS and
  relative SDFWSR stopping rules, \cite{vats:knud:2018} draw
  correspondence between ESS and a version of the widely used
  Gelman-Rubin (GR) diagnostic presented in the next section.

\subsection{Gelman-Rubin diagnostic}
\label{sec:gelm}
The GR diagnostic appears to be the most popular method for
assessing samples obtained from running MCMC algorithms. The GR
diagnostic relies on multiple chains
$\{X_{i 0}, X_{i 1},\dots,X_{i n-1}\}, i=1,\dots,m$ starting at
initial points that are drawn from a density that is over-dispersed
with respect to the target density $\pi$. \cite{gelm:rubi:1992}
describe methods of creating an initial distribution, although in
practice, these initial points are usually chosen in some ad hoc
way. Using parallel chains, \cite{gelm:rubi:1992} construct two
estimators of the variance of $X$ where $X \sim \pi$, namely, the
within-chain variance estimate,
$W = \sum\limits_{i=1}^m\sum\limits_{j=0}^{n-1}(X_{ij} -
\bar{X_{i\cdot}})^2/(m(n-1))$, and the pooled variance estimate
$\hat{V} = ((n-1)/n)W + B/n$ where
$B/n = \sum\limits_{i=1}^m(\bar{X_{i\cdot}} -
\bar{X_{\cdot\cdot}})^2/(m-1)$ is the between-chain variance estimate,
and $\bar{X_{i\cdot}}$ and $\bar{X_{\cdot\cdot}}$ are the $i^{th}$
chain mean and the overall mean respectively, $i = 1, 2,\dots,
m$. Finally, \cite{gelm:rubi:1992} compare the ratio of these two
estimators to one. In particular, they calculate the potential scale
reduction factor (PSRF) defined by
\begin{equation}\label{eq:psrf}
  \hat{R}=\frac{\hat{V}}{W},
\end{equation}
and compare it to one.

\cite{gelm:rubi:1992} argue that since the chains are started from an
over-dispersed initial distribution, in finite samples, the numerator in
(\ref{eq:psrf}) overestimates the target variance whereas the
denominator underestimates it, making $\hat{R}$ larger than
$1$. Simulation is stopped when $\hat{R}$ is sufficiently close to
one. The cutoff value 1.1 is generally used by MCMC practitioners, as
recommended by \cite{gelm:carl:ster:duns:veht:rubi:2014}. Recently,
\cite{vats:knud:2018} propose a modified GR statistic where the
between-chain variance $(B/n)$ is replaced with a particular batch
means estimator of the asymptotic variance for the Monte Carlo
averages $\bar{X}_n$. This modified definition allows for a connection
with ESS and, more importantly, computation of the GR diagnostic based
on a single
chain. 
We would like to point out that the expression of $\hat{R}$ given in
(\ref{eq:psrf}), although widely used in practice, differs slightly
from the original definition given by \cite{gelm:rubi:1992}.

\cite{broo:gelm:1998} propose the multivariate PSRF
(MPSRF) to diagnose convergence in the multivariate case. It is denoted by $\hat{R}_{p}$ and
is given by,
\begin{equation}\label{eq:mpsrf}
\hat{R}_{p} = \max_{a} \frac{a^{T}\widehat{V^\ast}a}{a^{T}W^\ast a} = \frac{n-1}{n} +
\bigg(1+\frac{1}{m}\bigg)~\lambda_{1},
\end{equation}
where $\widehat{V^\ast}$ is the pooled covariance matrix, $W^\ast$ is
the within-chain covariance matrix, $B^\ast$ is the between chain
covariance matrix and $\lambda_{1}$ is the largest eigenvalue of the
matrix $({W^\ast}^{-1}B^\ast)/n$. As in the univariate case,
simulation is stopped when $\hat{R}_{p} \approx
1$. \cite{pelt:venn:kask:2009} have proposed a visualization tool
based on linear discriminant analysis and discriminant component
analysis which can be used to complement the diagnostic tools proposed
by \cite{gelm:rubi:1992} and \cite{broo:gelm:1998}. The GR diagnostic
can be easily calculated, and is available in different statistical
packages including the CODA package \citep{R:coda} in R. To conclude
our discussion on the GR diagnostic, note that originally
\cite{gelm:rubi:1992} suggested running $m$ parallel chains, each of
length $2n$. Then discarding the first $n$ simulations, $\hat{R}$ is
computed based on the last $n$ iterations. This leads to the waste of
too many samples, and we do not recommend it.

\subsection{Two spectral density-based methods}
\label{sec:specd}

In this section, we discuss
two diagnostic tools based on asymptotic variance estimates of certain
statistics to check for convergence to
stationarity. \cite{gewe:1992} proposes a diagnostic tool based on the
assumption of existence of the spectral density of a related time
series. Indeed, for the estimation of $E_\pi g$, the asymptotic
variance of $\bar{g}_n$ is $S_g(0)$, the spectral density of
$\{g(X_n), n \ge 0 \}$ (treated as a time series) evaluated at
zero. After $n$ iterations of the Markov chain, let $\bar{g}_{n_A}$
and $\bar{g}_{n_B}$ be the time averages based on the first $n_A$ and
the last $n_B$ observations.  \cite{gewe:1992}'s statistic is the
difference $\bar{g}_{nA} - \bar{g}_{nB}$, normalized by its standard
error calculated using a nonparametric estimate of $S_g(0)$ for the
two parts of the Markov chain. Thus, \cite{gewe:1992}'s statistic is
\[ Z_n = \Big(\bar{g}_{n_A} - \bar{g}_{n_B}\Big) \Big/
  \sqrt{\widehat{S_g(0)}/n_A + \widehat{S_g(0)}/n_B}.
\]
\cite{gewe:1992}
suggests using $n_A = 0.1n$ and $n_B = 0.5n$. The $Z$ score is
calculated under the assumption of independence of the two parts of
the chain. Thus \cite{gewe:1992}'s convergence diagnostic is a $Z$
test of equality of means where autocorrelation in the samples is
taken into account while calculating the standard error.

\cite{heid:welc:1983} propose another method based on spectral
density estimates.  \cite{heid:welc:1983}'s diagnostic is based on
\[B_n(t) = \Big(\sum_{i=0}^{\lfloor nt\rfloor} g(X_i) - \lfloor
  nt\rfloor \bar{g}_n\Big) \Big/ \sqrt{n \widehat{S_g(0)}}.
\]
Assuming
that $\{B_n(t), 0 \le t \le 1\}$ is distributed
asymptotically as a Brownian bridge, the Cramer-von Mises statistic
$\int_0^1 B_n(t)^2 dt$ may be used to test the stationarity of the
Markov chain. The stationarity test is successively applied, first on
the whole chain, and then rejecting the first $10 \%, 20 \%, \dots $
and so forth of the samples until the test is passed or $50 \%$ of the samples have been
rejected. Both of these two spectral density-based tools presented
here are implemented in the CODA package \citep{R:coda}. These are
univariate diagnostics although \cite{cowl:carl:1996} mention that for
\cite{gewe:1992}'s statistic, $g$ may be taken to be $-2$ times the log
of the target density when $\mathcal{X} = \mathbb{R}^d$ for some
$d >1$. Finally, we would like to mention that the two spectral
density based methods mentioned here, just like the ESS and the GR
diagnostic, assume the existence of a Markov chain CLT (\ref{eq:clt}),
emphasizing the importance of the theoretical analysis discussed in
Section~\ref{sec:hone}.

\subsection{Raftery-Lewis diagnostic}
\label{sec:raftlew}

Suppose the goal is to estimate a quantile of $g(X)$, that is, to
estimate $u$ such that $P_\pi(g(X) \le u) =q$ for some prespecified
$q$. \cite{raft:lewi:1992} propose a method for calculating an
appropriate burn-in. They also discuss choosing a run length so that
the resulting probability estimate lies in
$[q- \epsilon, q + \epsilon]$ with probability $(1- \alpha)$. Thus the
required accuracy $\epsilon$ is achieved with probability
$(1- \alpha)$. \cite{raft:lewi:1992} consider the binary process
$W_n \equiv I(g(X_n) \le u), n \ge 0$. Although, in general,
$\{W_n\}_{n \ge 0}$ itself is not a Markov chain,
\cite{raft:lewi:1992} assume that for sufficiently large $k$, the
subsequence $\{W_{nk}\}_{n \ge 0}$ is approximately a Markov
chain. They discuss a method for choosing $k$ using model selection
techniques. The transition probability $P(W_{nk} = j | W_{(n-1)k} =i)$
is estimated by the usual estimator
\[
  \frac{\sum_{l = 1}^{n} I(W_{lk} = j, W_{(l-1)k} =i)}{\sum_{l = 1}^{n} I(W_{lk} =i)},
\]
for $i, j =0,1$. Here, $I(\cdot)$ is the indicator
function. Using a straightforward eigenvalue analysis of the
two-state empirical transition matrix of $\{W_{nk}\}_{n \ge 0}$,
\cite{raft:lewi:1992} provide an estimate of the burn-in. Using a CLT
for $ \sum_{j=0}^{n-1} W_{jk}/n$, they also give a stopping rule to
achieve the desired level of accuracy.

To implement this univariate method an initial number $n_{\text{min}}$
of iterations is used, and then it is determined if any additional runs are
required using the above techniques. The value
$n_{\text{min}} = \{\Phi^{-1}(1 -\alpha/2)\}^2 q(1-q)/\epsilon^2$ is
based on the standard asymptotic sample size calculation for Bernoulli
$(q)$ population. Since the diagnostic depends on the $q$ values, the
method should be repeated for different quantiles and the largest
among these burn-in estimates can be used. \cite{raft:lewi:1992}'s
diagnostic is available in the CODA package \citep{R:coda}.

\subsection{Kernel density-based methods}
\label{sec:kern}
There are MCMC diagnostics which compute distance between the kernel
density estimates of two chains or two parts of a single chain and
conclude convergence when the distance is close to zero. Unlike the
widely used GR diagnostic \citep{gelm:rubi:1992} which is based on
comparison of some summary moments of MCMC chains, these tools are
intended to assess the convergence of the whole
distributions. \cite{yu:1994} and \cite{boon:merr:krac:2014} estimate
the $L^1$ distance and Hellinger distance between the kernel density
estimates respectively. More recently, \cite{dixi:roy:2017} use the
symmetric Kullback Leibler (KL) divergence to produce two diagnostic
tools based on kernel density estimates of the chains. Below, we
briefly describe the method of \cite{dixi:roy:2017}.

Let $\{X_{ij}: i=1,2; j=1,2,\dots, n\}$ be the $n$ observations
obtained from each of the two Markov chains initialized from two points
well separated with respect to the target density $\pi$. The adaptive
kernel density estimates of observations obtained from the two chains
are denoted by $p_{1n}$ and $p_{2n}$ respectively. The KL divergence
between $p_{in}$ and $p_{jn}$ is denoted by
$KL(p_{in}|p_{jn}), i\neq j,\; i,j=1, 2$, that is,
\[
  KL(p_{in}|p_{jn}) = \int_{\mathcal{X}} p_{in}(x) \log\frac{p_{in}(x)}{p_{jn}(x)}  dx.
\]
\cite{dixi:roy:2017} find the Monte Carlo estimates of
$KL(p_{in}|p_{jn})$ using samples simulated from $p_{in}$ using the
technique proposed by \citet[][Sec 6.4.1]{silv:1986}. They use the
estimated symmetric KL divergence
($[KL(p_{1n}|p_{2n}) + KL(p_{2n}|p_{1n})]/2$) between $p_{1n}$ and
$p_{2n}$ to assess convergence where a testing of hypothesis framework
is used to determine the cutoff points.  The hypotheses are chosen
such that the type 1 error is concluding that the Markov chains have
converged when in fact they have not. The cutoff points for the
symmetric KL divergence are selected to ensure that the probability of
type 1 error is below some level say, 0.05. In case of multiple
($m > 2$) chains, the maximum among ${m \choose 2}$ estimated
symmetric KL divergences (referred to as Tool 1) is used to diagnose
MCMC convergence.  Finally, for multivariate examples---that is, when
$\mathcal{X} = \mathbb{R}^d$ for some $d >1$---although multivariate
Tool 1 can be used, in higher dimensions when kernel density
estimation is not reliable, \cite{dixi:roy:2017} recommend assessing
convergence marginally, i.e. one variable at a time, where appropriate
cutoff points are found by adjusting the level of significance using
Bonferroni's correction for multiple comparison.

For multimodal target distributions, if all chains are stuck at the
same mode, then empirical convergence diagnostics based solely on MCMC
samples may falsely treat the target density as unimodal and are prone
to failure. In such situations, \cite{dixi:roy:2017} propose another
tool (Tool 2) that makes use of the KL divergence between the kernel
density estimate of MCMC samples and the target density (generally
known up to the unknown normalizing constant) to detect
divergence. In particular, let $\pi(x) = f(x)/c$, where
$c = \int_{\mathcal{X}} f(x) dx$ is the unknown normalizing
constant. \cite{dixi:roy:2017}'s Tool 2 is given by
\begin{equation}
  \label{eq:tool2}
  T_2^\ast = \dfrac{\mid \hat{c} - c^\ast \mid}{c^\ast},
\end{equation}
where $\hat{c}$ is a Monte Carlo estimate, as described in section 3.3 of
\cite{dixi:roy:2017}, of the unknown normalizing
constant ($c$), based on the KL divergence between the adaptive kernel
density estimate of the chain and $\pi$, and $c^\ast$ is an estimate of $c$ obtained by
numerical integration. \cite{dixi:roy:2017} discuss that $T_2^\ast$
can be interpreted as the percentage of the target distribution not
yet captured by the Markov chain. Using this interpretation, they
advocate that if $T_2^\ast > 0.05$, then the Markov chain has not yet
captured the target distribution adequately. Since (\ref{eq:tool2}) involves
numerical integration, it cannot be used in high-dimensional examples.

{{\bf A visualization tool:}} \cite{dixi:roy:2017} propose a simple visualization tool to complement
their KL divergence diagnostic tool. This tool can be used for any
diagnostic method (including the GR diagnostic) based on multiple
chains started at distinct initial values, to investigate reasons
behind their divergence. Suppose $m (\ge 3)$ chains are run, and a
diagnostic tool has revealed that the $m$ chains have not mixed
adequately and thus the chains have not yet converged.  This
indication of divergence could be due to a variety of reasons. A
common reason for divergence is formation of clusters among multiple
chains.  \cite{dixi:roy:2017}'s visualization tool utilizes the tile
plot to display these clusters. As mentioned in
Section~\ref{sec:kern}, for $m$ chains, the KL divergence tool finds the estimated
symmetric KL divergence between each of the ${m \choose 2}$
combinations of chains and reports the maximum among them. In the visualization
tool, if the estimated symmetric KL divergence for a
particular combination is less than or equal to the cutoff value,
then the tool utilizes a gray tile to represent that the two chains
belong to the same cluster, or else it uses a black tile to
represent that the two chains belong to different clusters.

This visualization tool can also be used for
multivariate chains. In cases where the diagnostic tool for $d$
variate chains indicates divergence, for further investigation, the
user can choose a chain from each cluster and implement the
visualization tool marginally i.e. one variable at a time. This will
help the user identify which among the $d$ variables are responsible
for inadequate mixing among the $m$ multivariate chains.

\subsection{Graphical methods}
\label{sec:graph}

In addition to the visualization tool mentioned in
Section~\ref{sec:kern}, we now discuss some of the widely used
graphical methods for MCMC convergence diagnosis. The most common
graphical convergence diagnostic method is the trace plot. The trace
plot is a time series plot that shows the realizations of the Markov
chain at each iteration against the iteration numbers. This graphical
method is used to visualize how the Markov chain is moving around the
state space, that is, how well it is mixing. If the MCMC chain is
stuck in some part of the state space, the trace plots shows flat bits
indicating slow convergence. Such a trace plot is observed for an MH
chain if too many proposals are rejected consecutively. In contrast,
for an MH chain if too many proposals are accepted consecutively, then
trace plots may move slowly not exploring the rest of the state
space. Visible trends or changes in spread of the trace plot imply that
the stationarity has not been reached yet. It is often said that a
good trace plot should look like a hairy caterpillar. For an efficient
MCMC algorithm if the initial value is not in the high-density region,
the beginning of the trace plots shows back-to-back steps in one
direction. On the other hand, if the trace plot shows similar pattern
throughout, then there is no use in throwing burn-in samples.

Unlike iid sampling, MCMC algorithms result in correlated samples. The
lag-$k$ (sample) autocorrelation is defined to be the correlation
between the samples $k$ steps apart. The autocorrelation plot shows
values of the lag-$k$ autocorrelation function (ACF) against
increasing $k$ values. For fast-mixing Markov chains, lag-$k$
autocorrelation values drop down to (practically) zero quickly as $k$
increases. On the other hand, high lag-$k$ autocorrelation values for
larger $k$ indicate the presence of a high degree of correlation and slow
mixing of the Markov chain. Generally, in order to get precise Monte Carlo
estimates, Markov chains need to be run a large multiple of the amount of time
it takes the ACF to be practically zero.

Another graphical method used in practice is the running mean plot
although its use has faced criticism \citep{geye:2011}. The running
mean plot shows the Monte Carlo (time average) estimates against the
iterations. This line plot should stabilize to a fixed number as
iteration increases, but non-convergence of the plot
indicates that the simulation cannot be stopped yet. While the trace
plot is used to diagnose a Markov chain's convergence to stationarity,
the running mean plot is used to decide stopping times.

In the multivariate case, individual trace, autocorrelation and
running mean plots are generally made based on realizations of each
marginal chain. Thus the correlations that may be present among
different components are not visualized through these plots. In
multivariate settings, investigating correlation across different
variables is required to check for the presence of high
cross-correlation \citep{cowl:carl:1996}.

\section{Examples}
\label{sec:exam}
In this section, we use three detailed examples to illustrate the
convergence diagnostics presented in Section~\ref{sec:diag}. Using
these examples, we also demonstrate that empirical convergence
diagnostic tools may give false indication of convergence to
stationarity as well as convergence of Monte Carlo estimates. 

\subsection{An exponential target distribution}
\begin{figure*}
  \includegraphics[width=\linewidth]{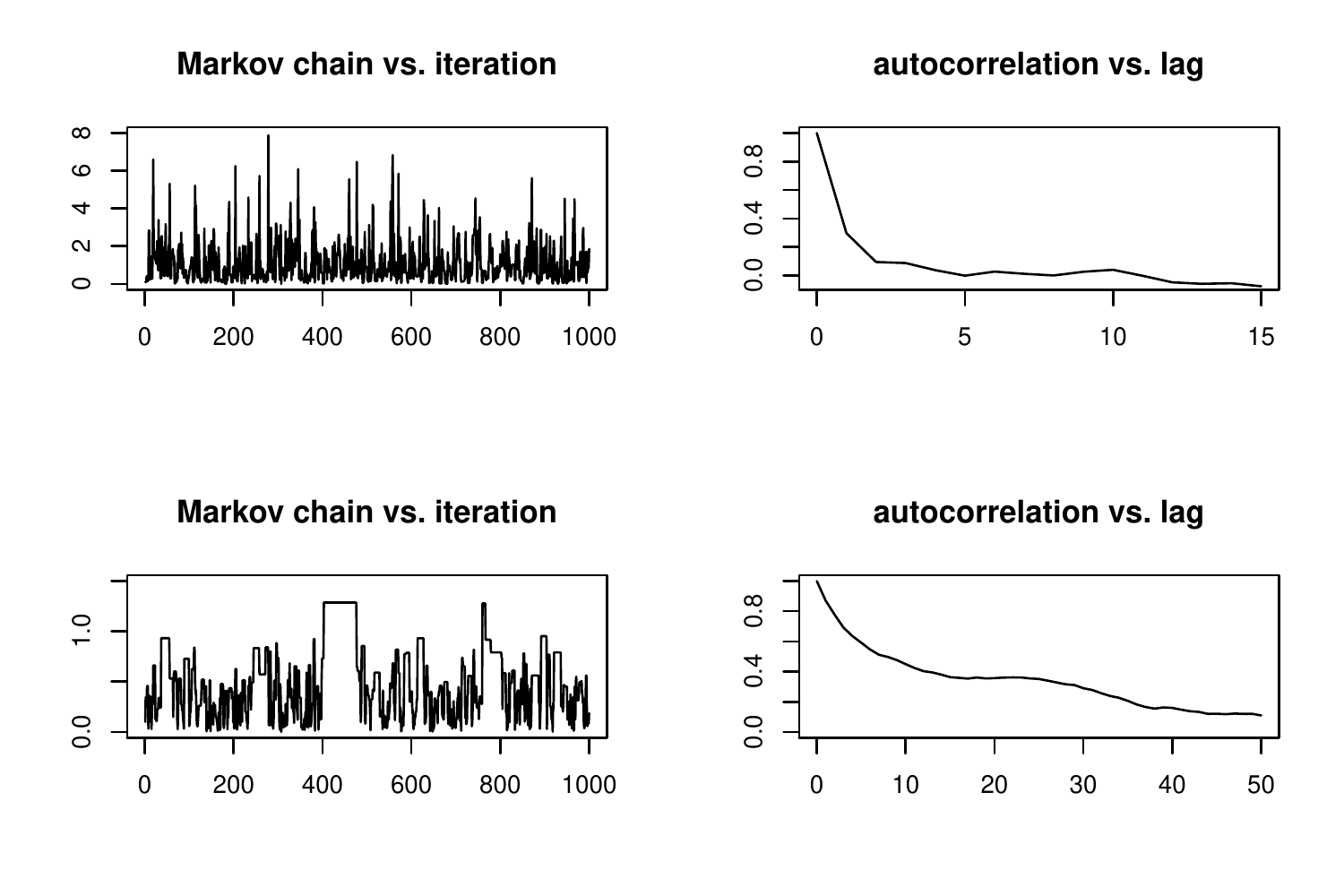}
  \caption{Trace (left panels) and autocorrelation function (right
    panels) plots of the independence Metropolis chains (top row,
    $\theta =0.5$; bottom row, $\theta =5$) for the exponential target
    example. The presence of frequent flat bits in the trace plot and high autocorrelation values indicate
  slow mixing of the Markov chain with $\theta =5$.}
 \label{fig:exp:tr.acf}
  \end{figure*}

Let the target distribution be Exp(1), that is,
$\pi(x) = \exp(-x), \; x>0$. We consider an independence Metropolis
sampler with Exp($\theta$) proposal, that is, the proposal density is
$q(x) = \theta \exp(-\theta x), \; x>0$. We study the independence
chain corresponding to two values of $\theta$, namely, $\theta =0.5$
and $\theta =5$. Using this example, we illustrate the honest
choices of burn-in and stopping time described in
Section~\ref{sec:hone} as well as several other diagnostic tools. It
turns out that, even in this unimodal example, some empirical
diagnostics may lead to premature termination of simulation. We first
consider some graphical diagnostics for Markov chain
convergence. Since the target density is a strictly decreasing
function on $(0, \infty)$, a small value may serve as a reasonable
starting value.  We run the Markov chains for 1,000 iterations
initialized at $X_0 =0.1$. Figure~\ref{fig:exp:tr.acf} shows the trace
plots and autocorrelation plots of the Markov chain samples. From the
trace plots we see that while the first chain ($\theta =0.5$) mixes
well, the second chain exhibits several flat bits and suffers from
slow mixing. Thus from the trace plots, we see that there is no need
for burn-in for $\theta =0.5$, that is, $X_0 = 0.1$ seems to be a
reasonable starting value. On the other hand, for $\theta =5$, the
chain can be run longer to find an appropriate burn-in. This is also
corroborated by the autocorrelation plots. When $\theta =0.5$,
autocorrelation is almost negligible after lag 4. On the other hand,
for $\theta =5$, there is significant autocorrelation even after lag
50. Next, using the CODA package \citep{R:coda}, we compute
\cite{gewe:1992}'s and \cite{heid:welc:1983}'s convergence diagnostics
for the identity function $g(x)=x$. Using the default $n_A = 0.1n$ and
$n_B = 0.5n$, \cite{gewe:1992}'s Z scores for the $\theta =0.5$ and
$\theta =5$ chains are 0.733 and 0.605 respectively, failing to reject
the hypothesis of the equality of means from the beginning and end
parts of the chains. Similarly, both the chains pass
\cite{heid:welc:1983}'s test for stationarity. Next, we consider
the \cite{raft:lewi:1992} diagnostic. When the two samplers are run for
38,415 ($\lceil n_{min} \rceil$ corresponding to
$\epsilon= 0.005, \alpha=0.05$, and $q=0.5$) iterations, and
Raftery-Lewis diagnostic is applied for different $q$ values (0.1,
$\dots$, 0.9), the burn-in estimates for the $\theta =5$ chain are
larger than those for the $\theta =0.5$ chain, although the overall
maximum burn-in (981) is less than 1,000. Finally, we consider the
choice of honest burn-in. Since for $\theta < 1$,
$\pi(x)/q(x) = \theta^{-1} \exp(x(\theta -1)) \le \theta^{-1}$ for all
$x >0$, according to \cite{meng:twee:1996}, we know that
\[
  \frac{1}{2} \int_{\mathcal{X}} |f_n(x) - \pi (x)| dx \le (1 - \theta)^n,
\]
that is, an analytical upper bound to the TV norm can be obtained. Thus
for $\theta = 0.5$, if $n' = \lceil \log(0.01)/\log(0.5)\rceil = 7$,
then (\ref{eq:totvar}) holds. Thus $n' =7$ can be an honest burn-in
for the independence Metropolis chain with $\theta =0.5$. 
Note that, for $\theta < 1$, the independence chain is geometrically
ergodic; for $\theta = 1$, the chain produces iid draws from the
target; and for $\theta > 1$, by \cite{meng:twee:1996}, the independence
chain is {\it subgeometric}. As mentioned by
\cite{jone:hobe:2001}, when $\theta > 1$, the tail of the proposal
density is much lighter than that of the target, making it difficult
for the chain to move to larger values, and when it does move there,
it tends to get stuck.

\begin{figure*}
  \begin{minipage}[b]{0.5\linewidth}
    \includegraphics[width=\linewidth]{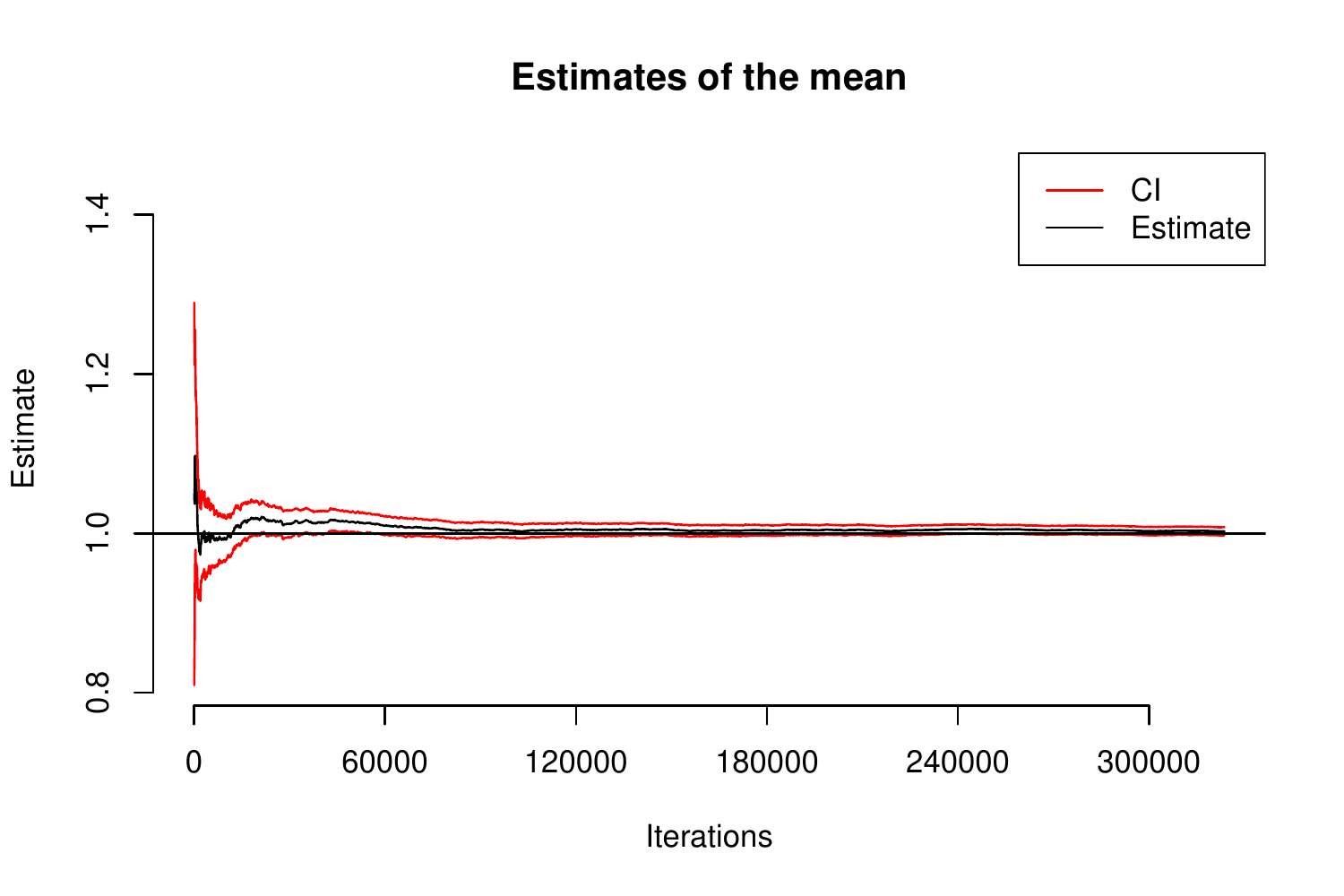}
  \end{minipage}
  \begin{minipage}[b]{0.5\linewidth}
    \includegraphics[width=\linewidth]{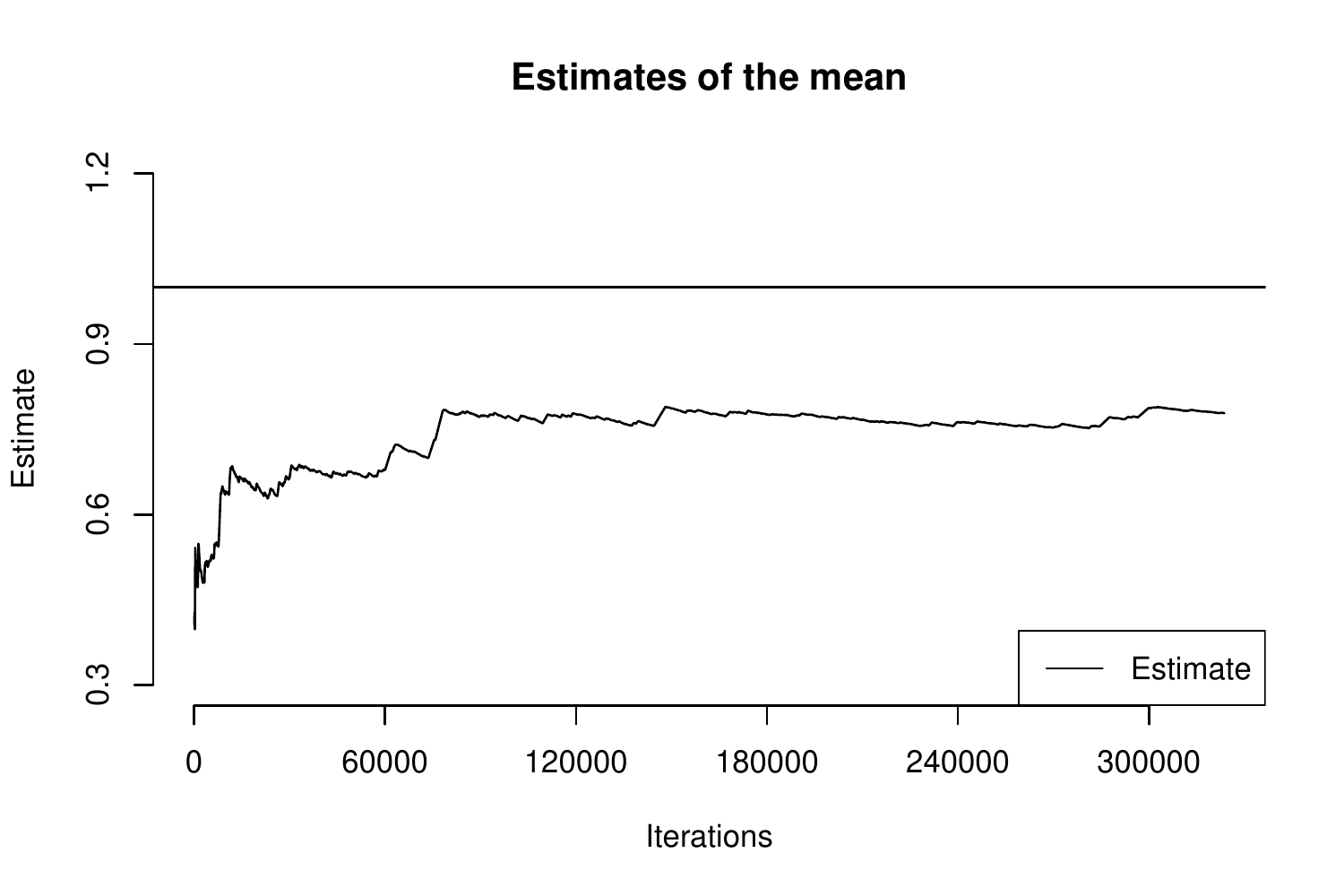}
  \end{minipage}
  \caption{The left plot shows the running estimates of the mean with
    confidence interval for $\theta =0.5$. Running mean plot for
    $\theta =5$ is given in the right panel. The horizontal line
    denotes the truth. The plot in the right panel reveals that even
    after 300,000 iterations, the Monte Carlo estimate for the chain
    with $\theta =5$ is far off from the truth.}
\label{fig:exp.run}
\end{figure*}

\begin{figure*}
  \begin{minipage}[b]{0.5\linewidth}
    \includegraphics[width=\linewidth]{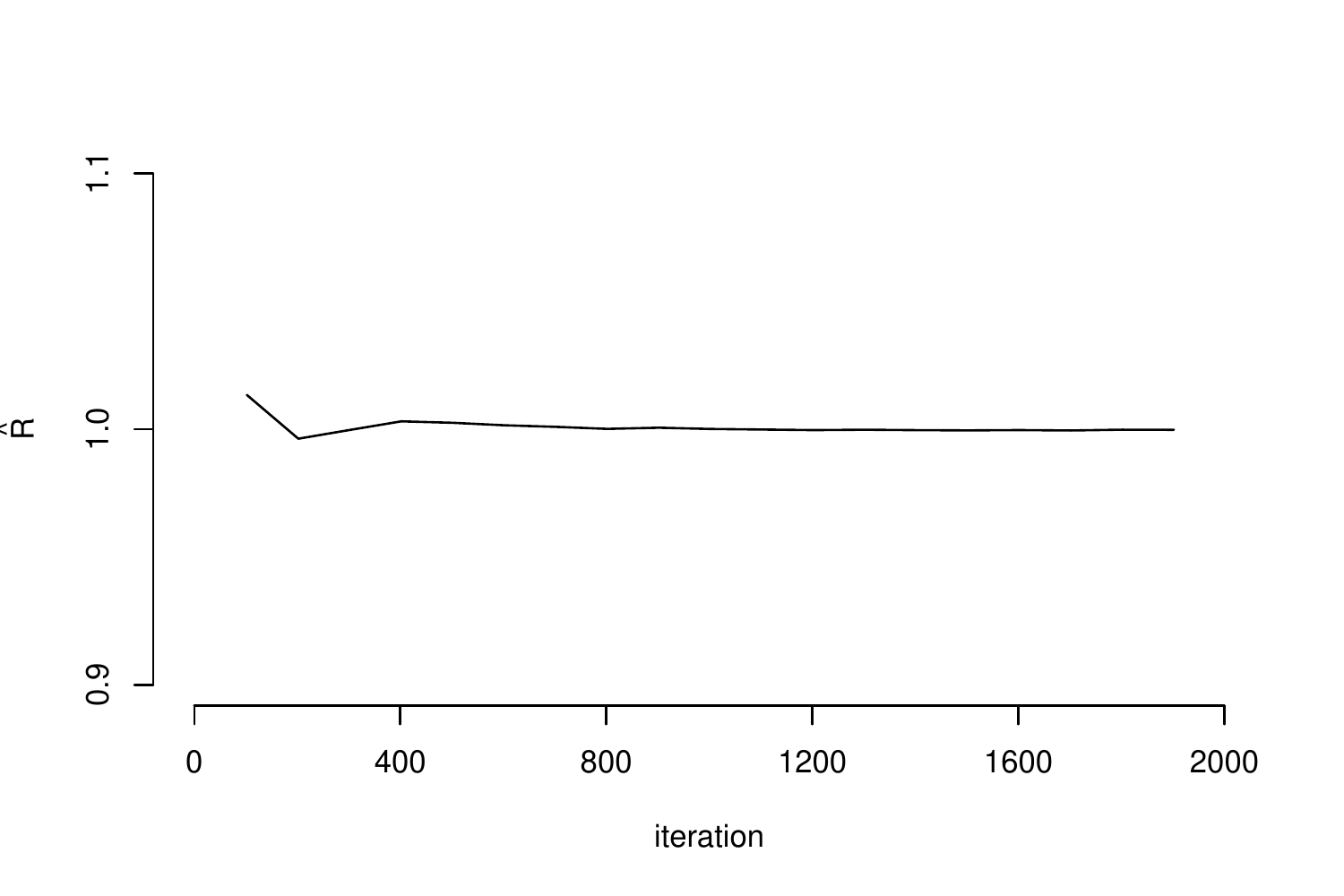}
  \end{minipage}
  \begin{minipage}[b]{0.5\linewidth}
    \includegraphics[width=\linewidth]{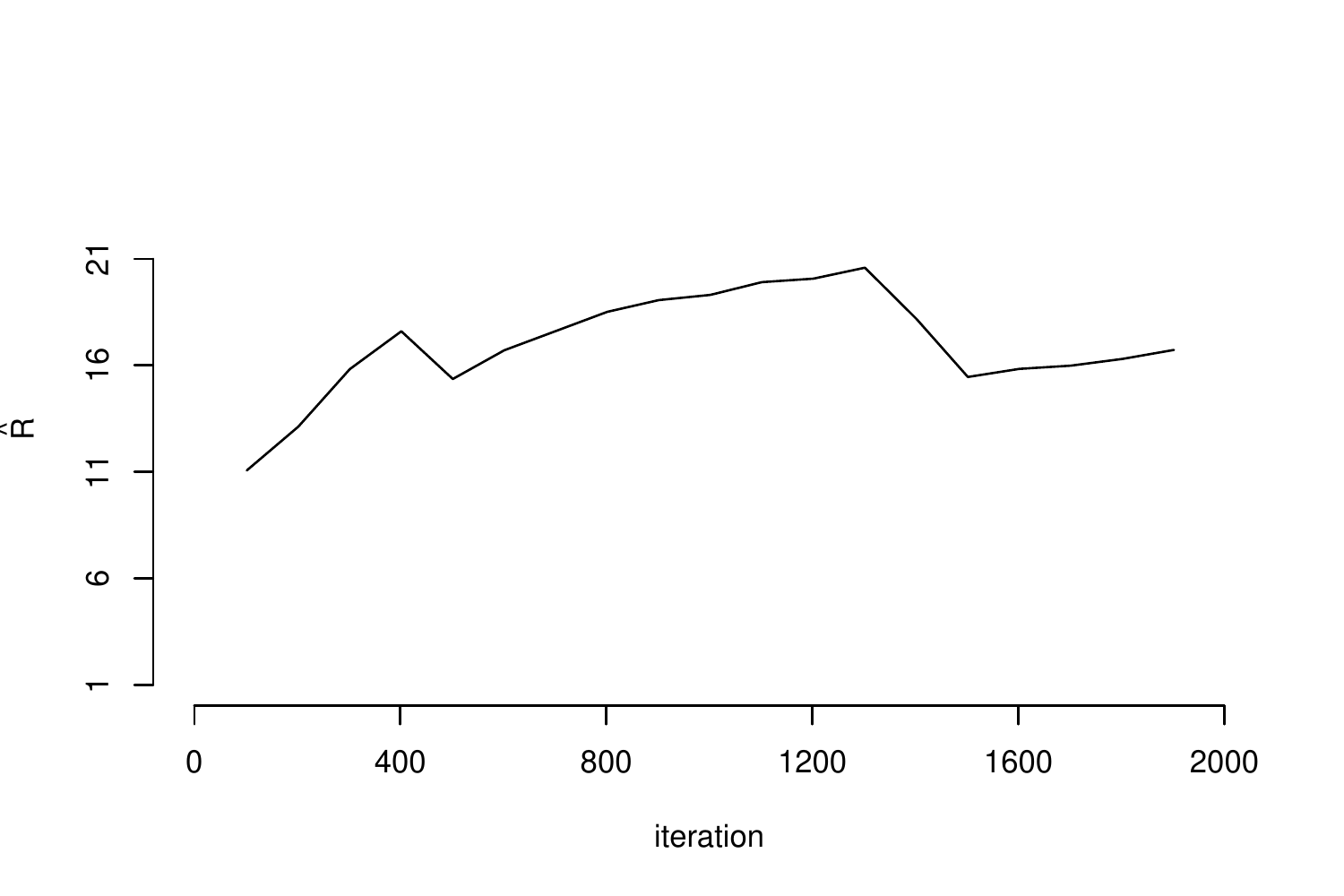}
  \end{minipage}
  \caption{Iterative $\hat{R}$ plot (from four parallel chains) for
    the independence chains (left plot $\theta =0.5$, right plot
    $\theta =5$). In the left plot, the PSRF reaches below the cutoff
    (1.1) before 100 iterations, leading to premature termination of
    the chain.}
\label{fig:exp.gr}
\end{figure*}

  Next, we consider stopping rules for estimation of the mean of the
  stationary distribution, that is, $E_\pi X = 1$. Based on a single
  chain, we apply the FWSR (\ref{eq:fwsr}) to determine the sample
  size for $\epsilon = 0.005$ and $\alpha =0.05$ (that is,
  $t_* = 1.96$).  For the independence Metropolis chain with
  $\theta =0.5$ starting at $X_{8} = 0.1545$, it takes $n^* = 323,693$
  iterations to achieve the cutoff $0.005$. The running estimates of
  the mean along with confidence intervals are given in the left panel
  of Figure~\ref{fig:exp.run}. We next run the independence Metropolis
  chain with $\theta =5$ for 323,700 iterations starting at
  $X_0 =0.1$. The corresponding running estimates are given in the
  right panel of Figure~\ref{fig:exp.run}. Since a Markov chain CLT is
  not available for the independence chain with $\theta >1$, we cannot
  compute asymptotic confidence intervals in this case. From the plot
  we see that the final estimate (0.778) is far off from the truth
  ($E_\pi X = 1$). Next, we consider ESS. The cut off value for ESS
  mentioned in (\ref{eq:messcut}) with $\varepsilon = 2*0.005=0.01$ is
  153,658. The ESS for the two chains are 163,955 and 1,166,
  respectively which again shows the presence of large correlation
  among the MCMC samples for $\theta =5$. We use the R package mcmcse
  \citep{R:mcmcse} for computing ESS. Finally, we consider the GR
  diagnostic. We run four parallel chains for 2,000 iterations
  starting at 0.1, 1, 2, and 3, respectively each with both
  $\theta =0.5$ and $\theta =5$. We calculate \cite{gelm:rubi:1992}'s
  PSRF (\ref{eq:psrf}) based on these chains. The plots of iterative
  $\hat{R}$ at the increment of every 100 iterations are given in
  Figure~\ref{fig:exp.gr}. We see that $\hat{R}$ for
  the chain with $\theta =0.5$ reaches below 1.1 in 100 iterations. On
  the other hand, the Monte Carlo estimate for $E_\pi X$ and its
  standard error based on first 100 iterations for the chain started
  at 0.1 are 1.109 and 0.111, respectively. Thus, GR diagnostic leads
  to premature termination of simulation and the inference drawn from
  the resulting samples can be unreliable. Finally, we note that $\hat{R}$ for the
  chains with $\theta =5$ takes large ($>16$) values even after 2,000
  iterations showing that simulation cannot be stopped yet in this
  case.

\subsection{A sixmodal target distribution}
\label{sec:sixmo}
\begin{figure}[h]
\begin{center}
\includegraphics[width=0.5\linewidth]{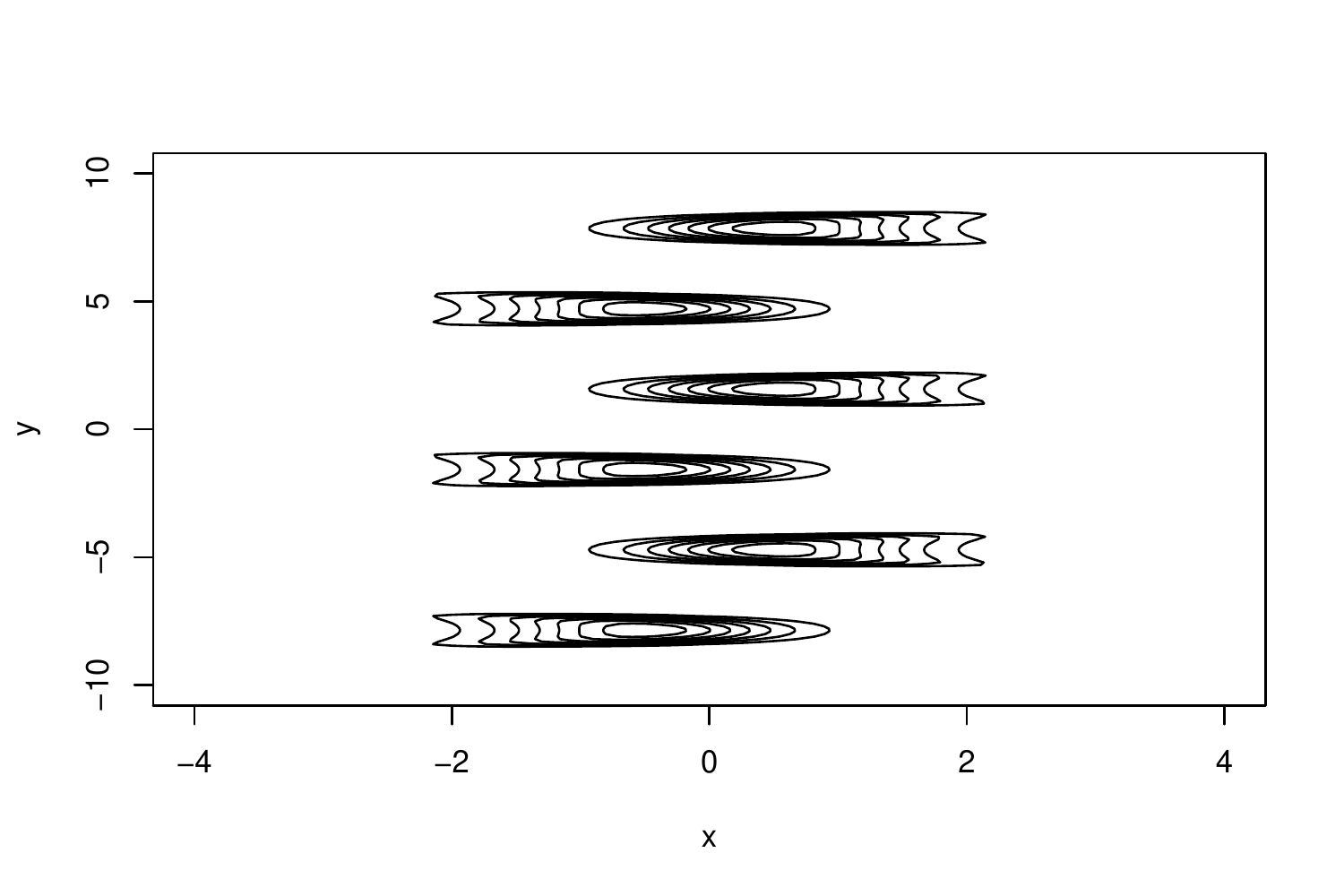}
\end{center}
\caption{Contour plot of the target distribution in the sixmodal example.}
\label{fig:six_joint}
\end{figure}

\begin{figure}[h]
\begin{center}
\includegraphics[width=0.5\linewidth]{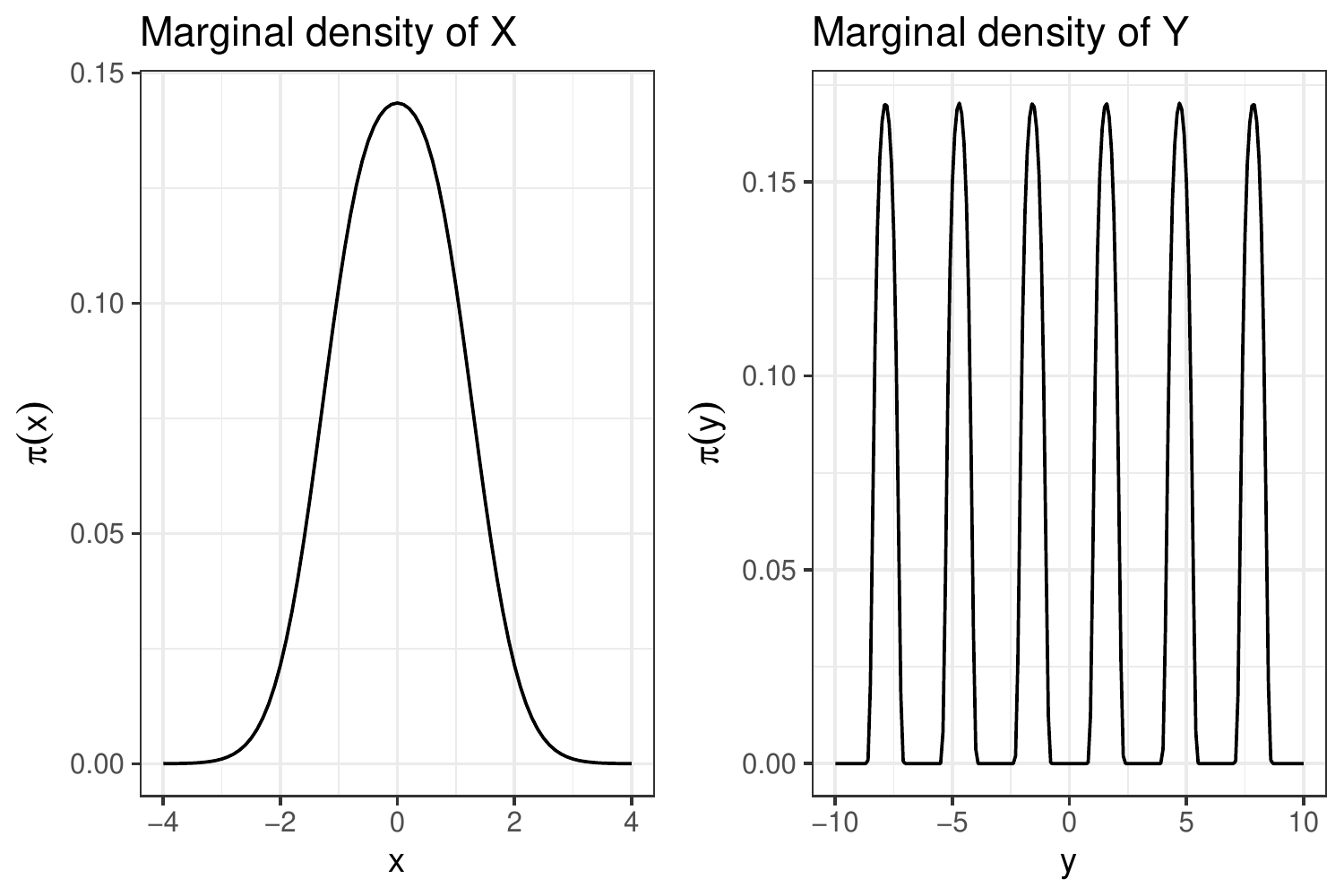}
\end{center}
\caption{Marginal densities of $X$ and $Y$ in the sixmodal example.}
\label{fig:six_x}
\end{figure}

This example is proposed by \cite{lem:chen:lavi:2009} where the
target density is as follows
\begin{equation}\label{eq:six_tar}
\pi(x, y)\propto\exp\bigg(\frac{-x^2}{2}\bigg)\exp\bigg(\frac{((\csc~ y)^5 - x)^2}{2}\bigg),\; -10 \le x, y \le 10.
\end{equation}
The contour plot of the target distribution (known up to the
normalizing constant) is given in Figure~\ref{fig:six_joint} and
marginal densities are plotted in Figure \ref{fig:six_x}. The plots of
the joint and marginal distributions clearly show that the target
distribution is multimodal in nature.

To draw MCMC samples from the target density (\ref{eq:six_tar}), we
use a Metropolis within Gibbs sampler in which $X$ is drawn first and
then $Y$. In this example, we consider only convergence to stationarity,
that is, we do not discuss stopping rules here. Through this example,
we illustrate that when an MCMC sampler is stuck in a local mode, the
empirical convergence diagnostic tools may give false indication of
convergence. [Empirical diagnostics may fail even when modes are not
well defined \citep{geye:thomp:1995}.] In order to illustrate the
diagnostic tools, as in \cite{dixi:roy:2017}, we consider two
cases.

\begin{figure*}
  \begin{minipage}[b]{0.5\linewidth}
    \includegraphics[width=\linewidth]{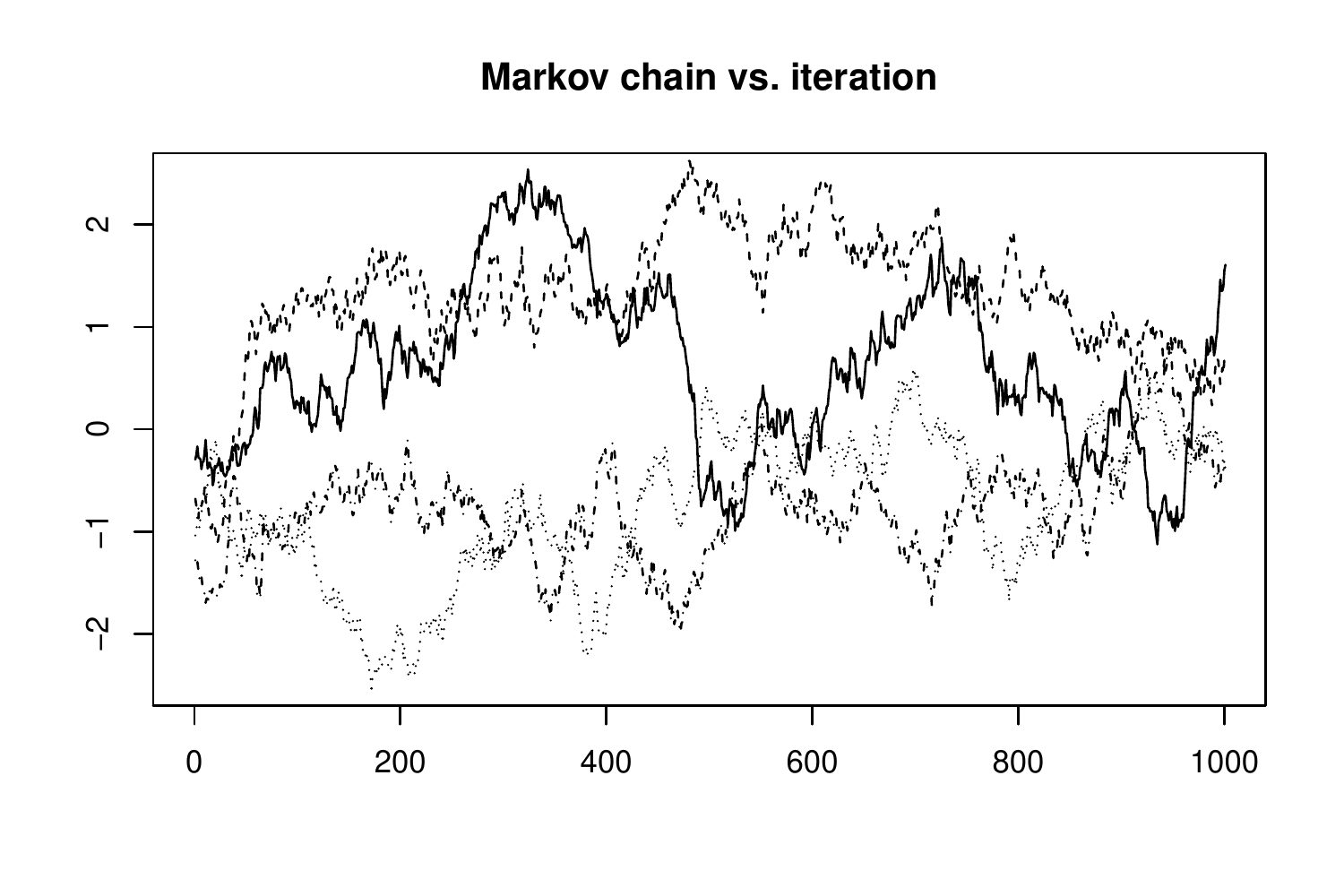}
  \end{minipage}
  \begin{minipage}[b]{0.5\linewidth}
    \includegraphics[width=\linewidth]{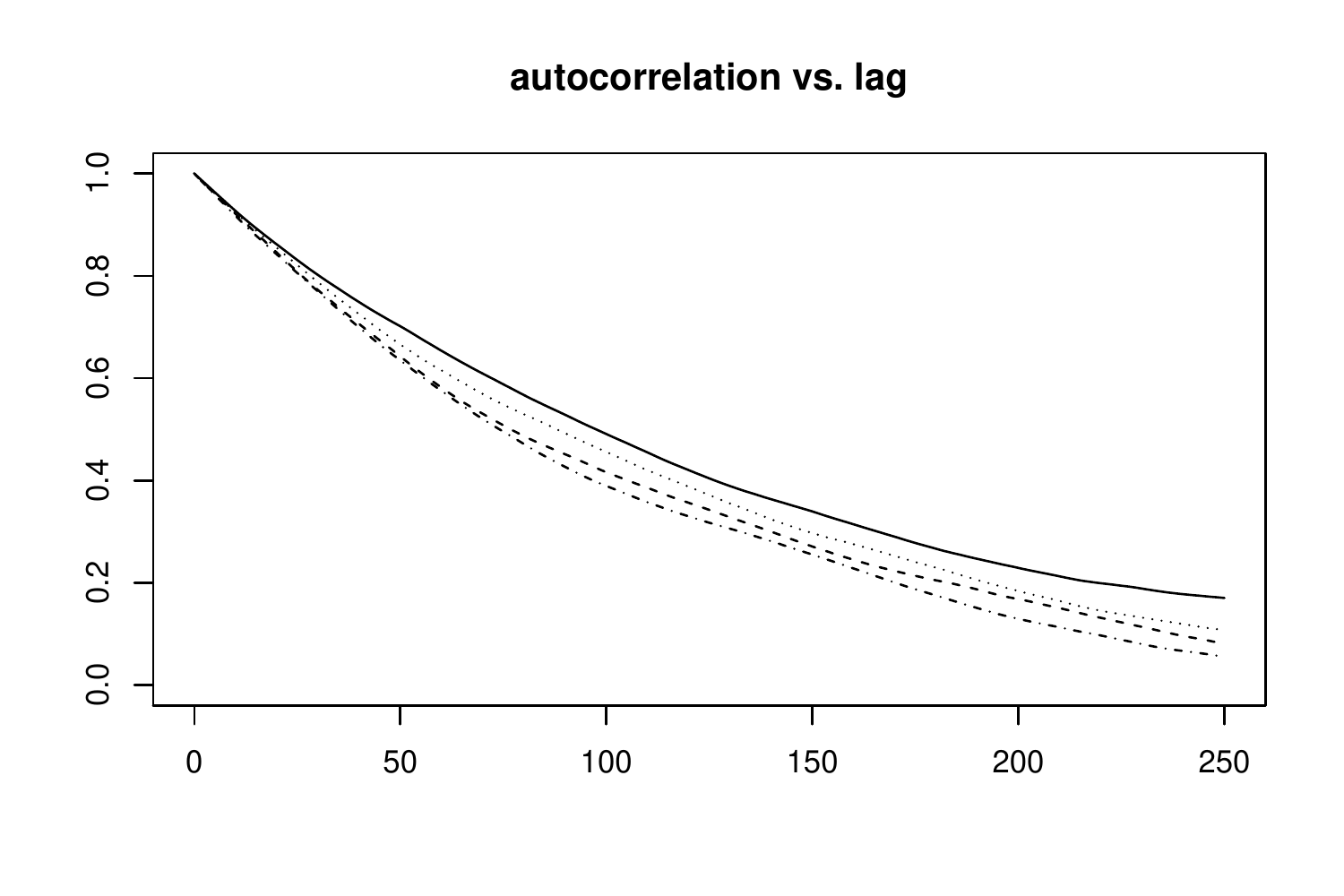}
  \end{minipage}
  \caption{Trace (left panel) and autocorrelation function (right panel) plots of the $X$ marginal of the four chains for the sixmodal example in Case 1.
}
\label{fig:tr.acf.x}
\end{figure*}

\begin{figure*}
  \begin{minipage}[b]{0.5\linewidth}
    \includegraphics[width=\linewidth]{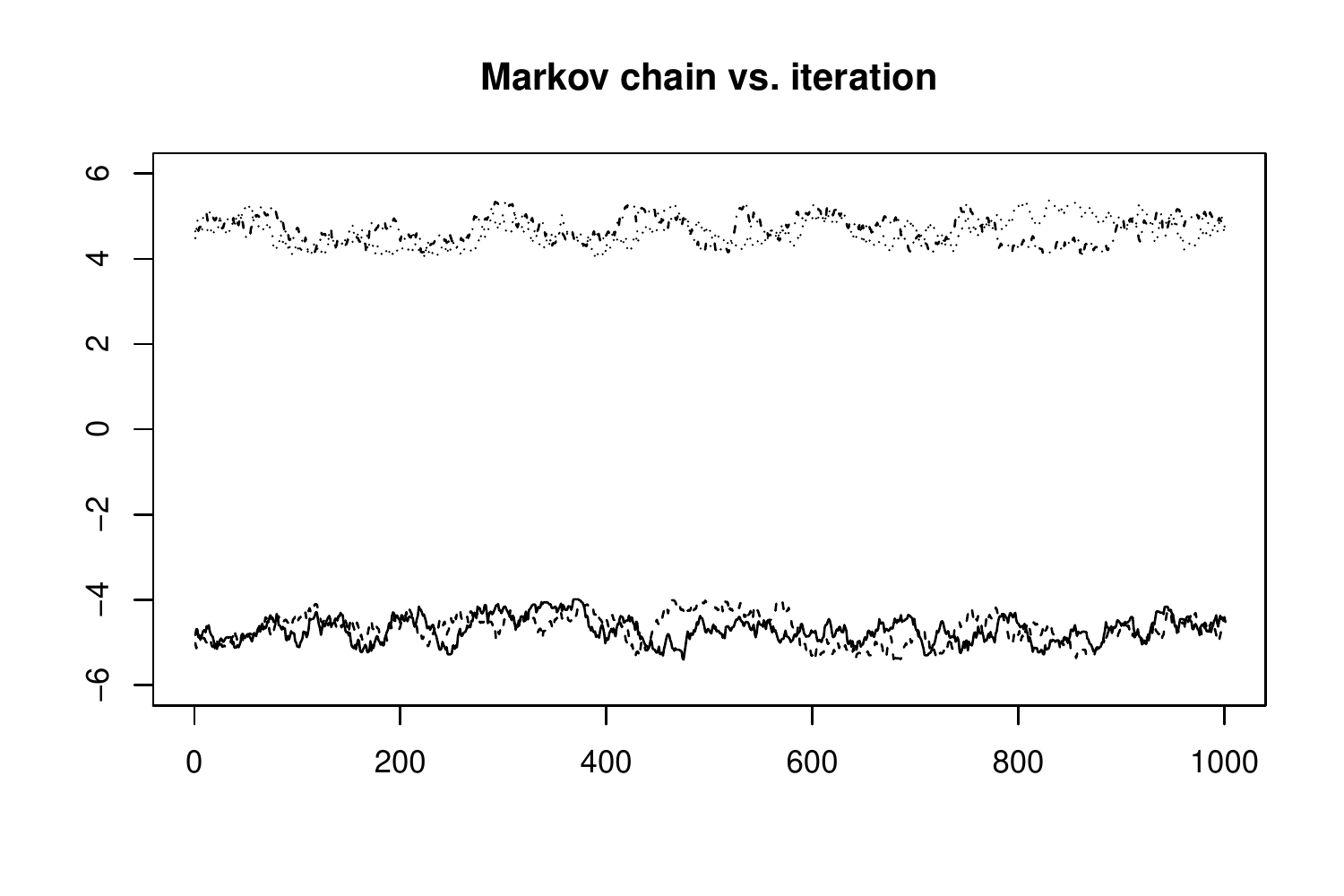}
  \end{minipage}
  \begin{minipage}[b]{0.5\linewidth}
    \includegraphics[width=\linewidth]{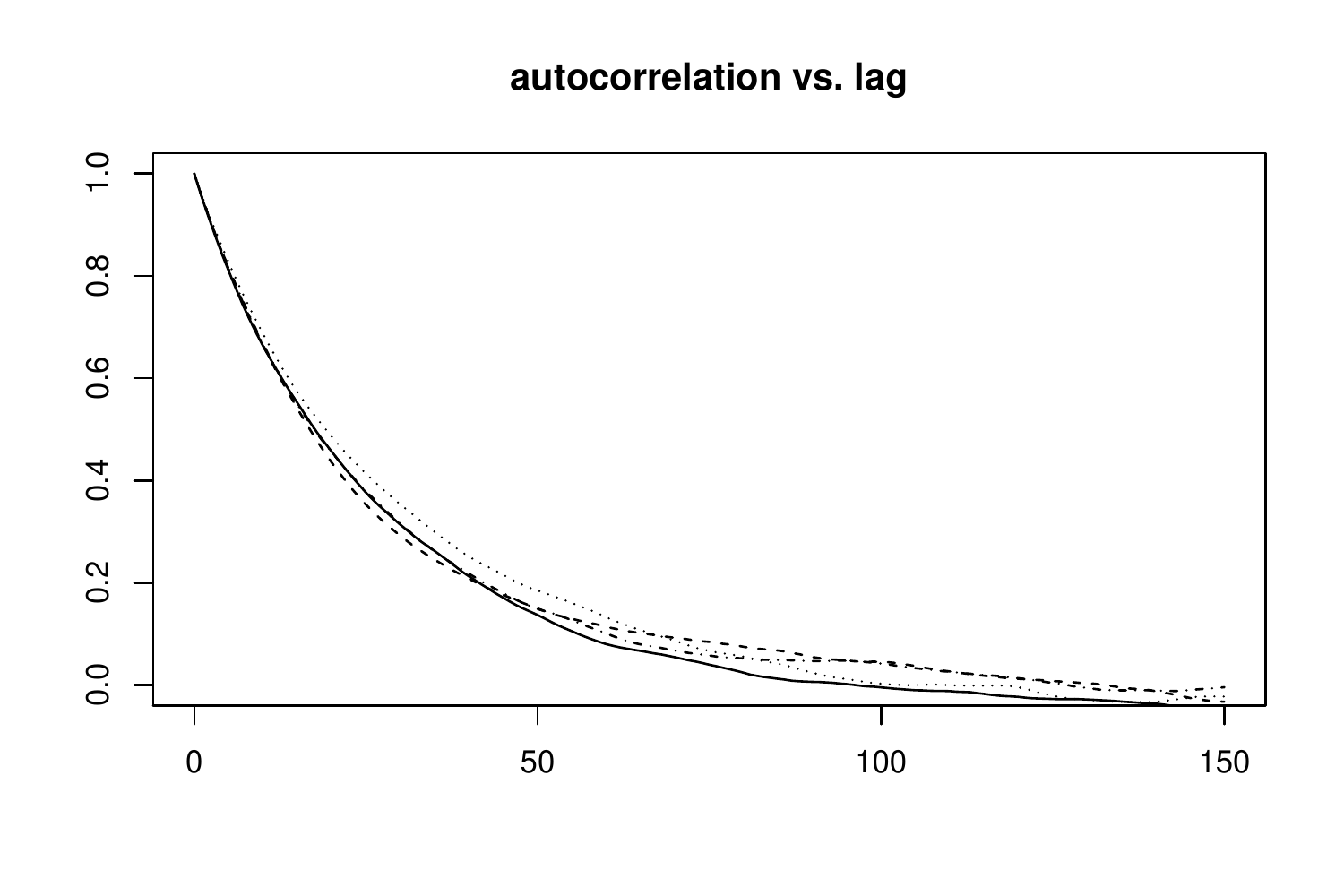}
  \end{minipage}
  \caption{Trace (left panel) and autocorrelation function (right panel) plots of the $Y$ marginal of the four chains for the sixmodal example in case 1.}
\label{fig:tr.acf.y}
\end{figure*}

\noindent \textbf{Case 1.} In this case, we run four chains wherein
two chains (chains 1 and 2) are started at a particular mode
while the remaining two chains (chains 3 and 4) are started at
some other mode. Each of the four chains is run for 30,000
iterations. Trace plots of the last one thousand iterations of the
four parallel $X$ and $Y$ marginal chains are given in the left panel
of Figures~\ref{fig:tr.acf.x} and \ref{fig:tr.acf.y}
respectively. Trace plots show the divergence of the Markov
chains. High ACF values can also be seen from the autocorrelation
plots of the marginal chains in Figures~\ref{fig:tr.acf.x} and
\ref{fig:tr.acf.y}.

\begin{figure}
\begin{center}
\includegraphics[width=3in]{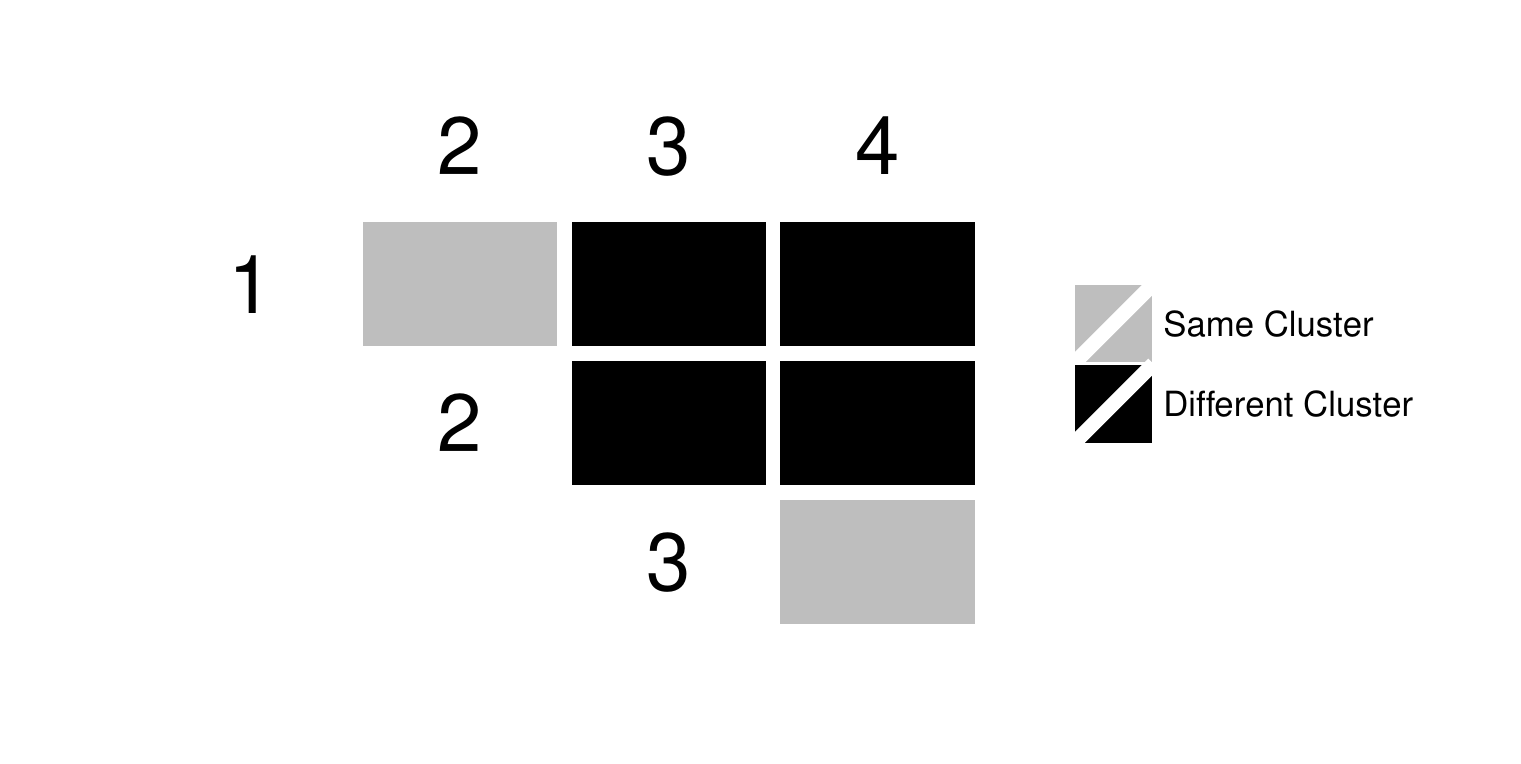}
\end{center}
\caption{\cite{dixi:roy:2017}'s tile plot in the Case 1 of the sixmodal example. The plot shows formation of two distinct clusters by the four chains.}
\label{fig:c1_viz}
\end{figure}

Next, we apply \cite{dixi:roy:2017}'s bivariate KL divergence Tool 1
on the joint chain. The maximum symmetric KL divergence among the six
pairs is 104.89 significantly larger than the cutoff value 0.06.
Finally, we use \cite{dixi:roy:2017}'s visualization tool to identify
clusters among the four chains. The result is given in Figure
\ref{fig:c1_viz} which shows that there are two clusters among the
four chains wherein chain 1 and chain 2 form one cluster, while chain
3 and chain 4 form another cluster.

\begin{figure*}
  \begin{minipage}[b]{0.5\linewidth}
    \includegraphics[width=\linewidth]{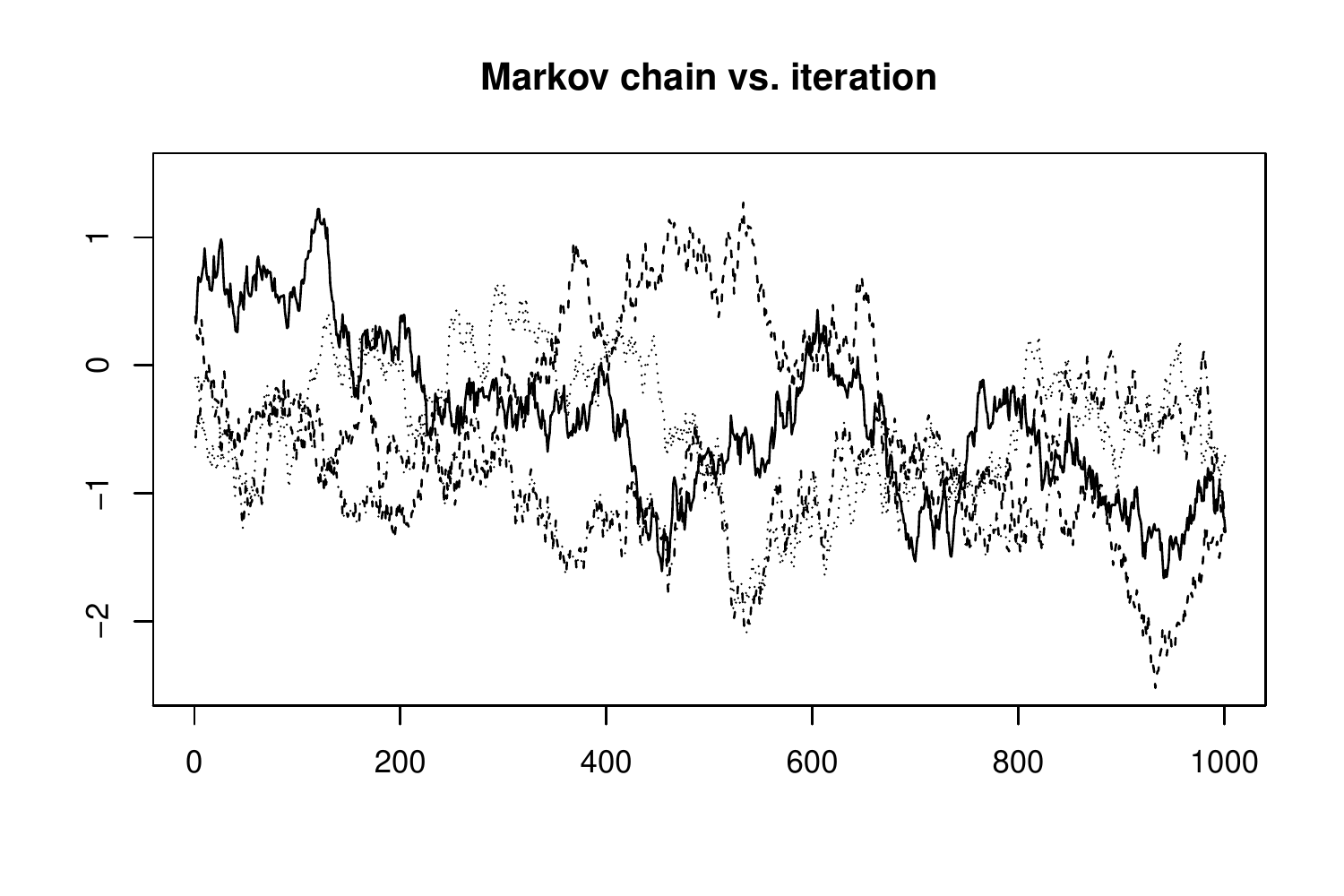}
  \end{minipage}
  \begin{minipage}[b]{0.5\linewidth}
    \includegraphics[width=\linewidth]{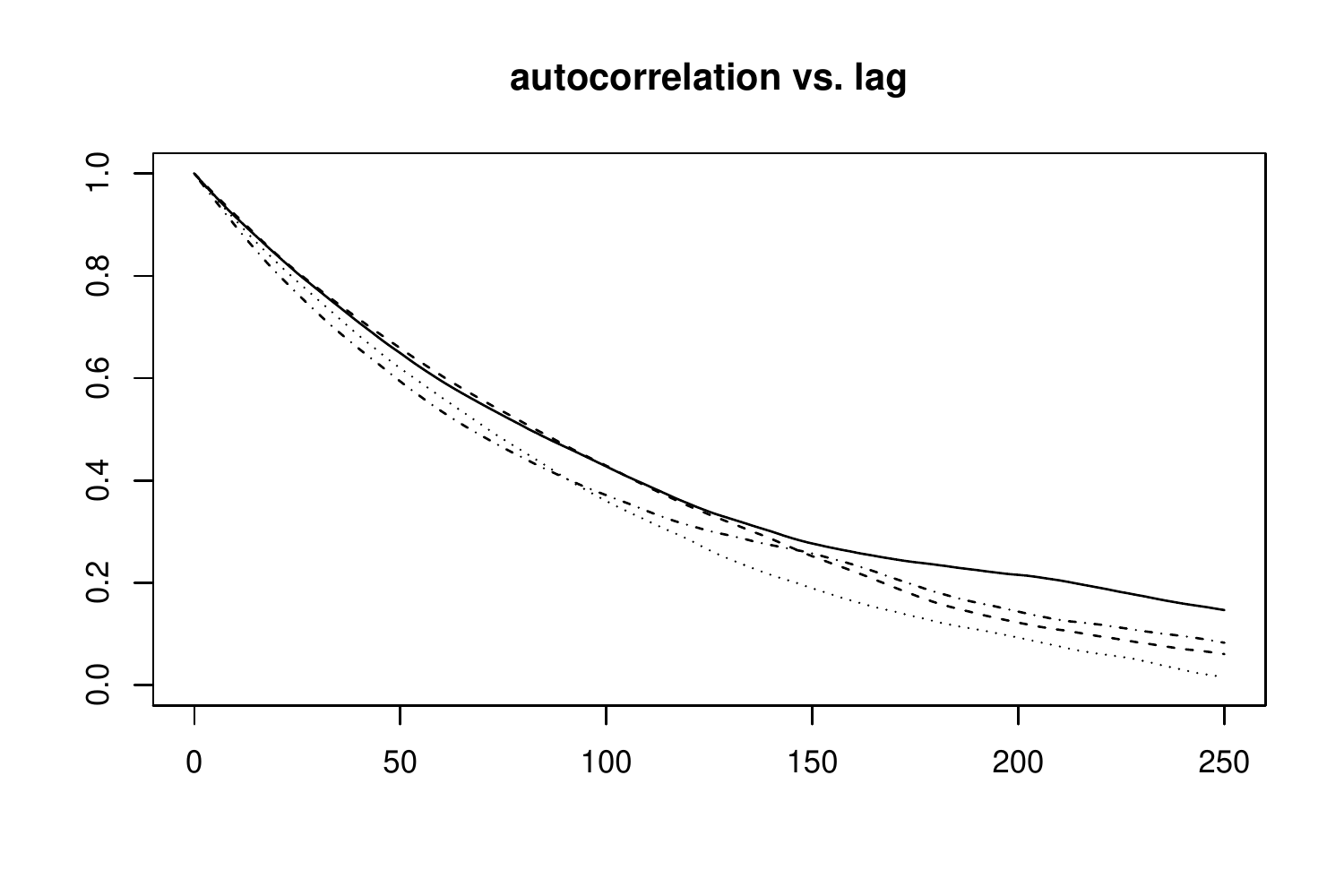}
  \end{minipage}
  \caption{Trace (left panel) and autocorrelation function (right panel) plots of the $X$ marginal of the four chains for the sixmodal example in case 2.
}
\label{fig:tr.acf.x2}
\end{figure*}

\begin{figure*}
  \begin{minipage}[b]{0.5\linewidth}
    \includegraphics[width=\linewidth]{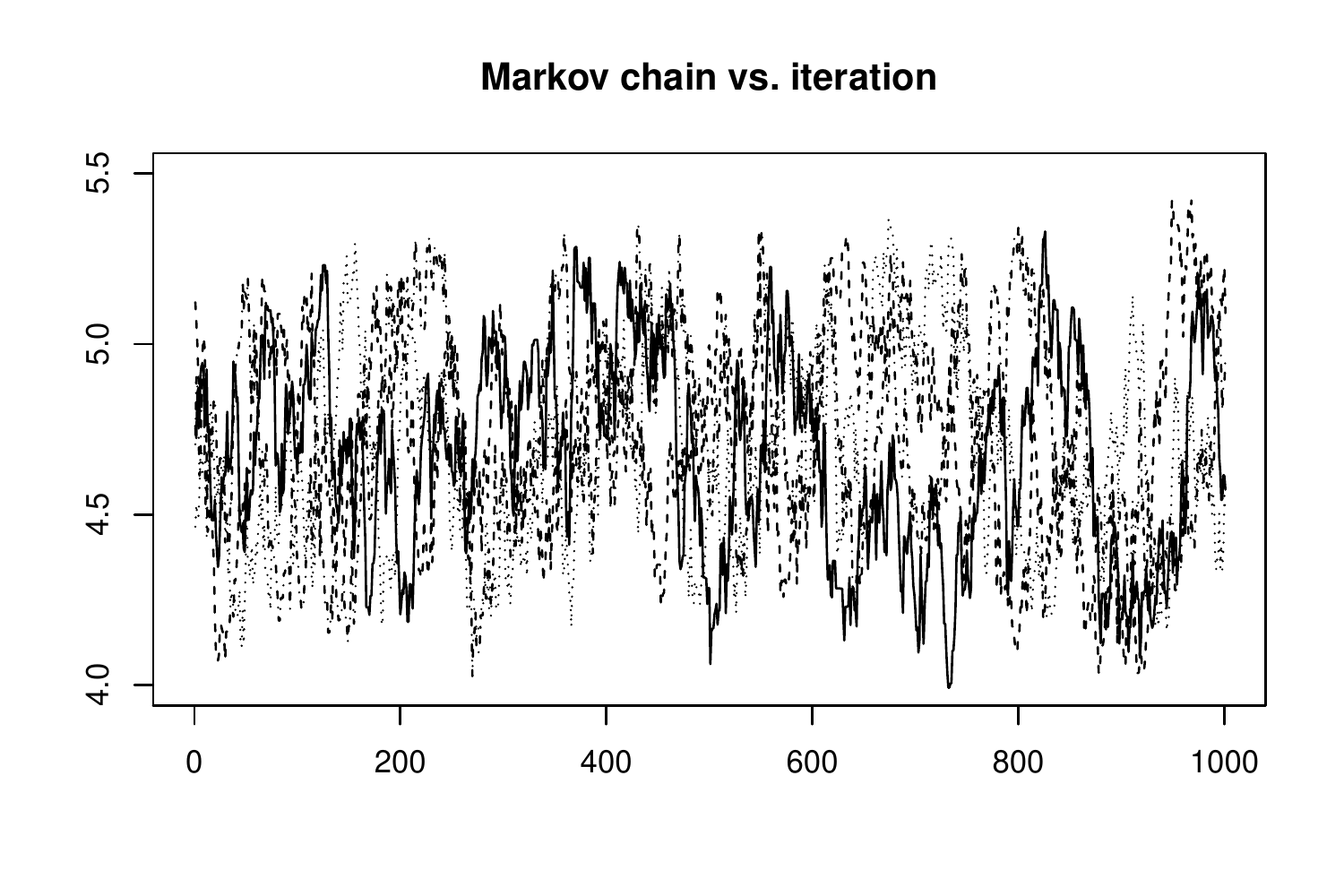}
  \end{minipage}
  \begin{minipage}[b]{0.5\linewidth}
    \includegraphics[width=\linewidth]{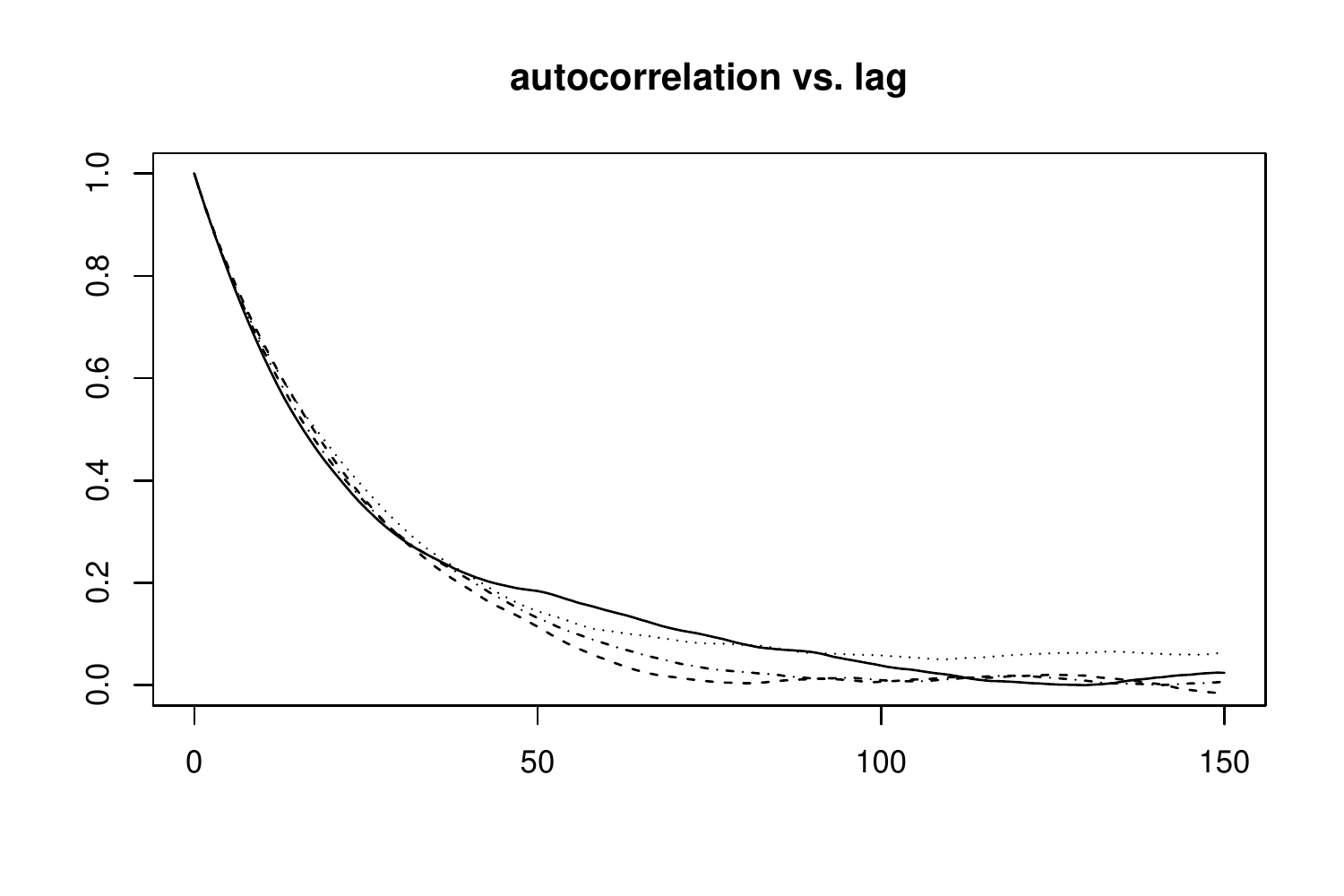}
  \end{minipage}
  \caption{Trace (left panel) and autocorrelation function (right panel) plots of the $Y$ marginal of the four chains for the sixmodal example in case 2.}
\label{fig:tr.acf.y2}
\end{figure*}

\noindent \textbf{Case 2:} In this case also we run four chains but all the chains are started at
the same local mode. As in Case 1, all four chains are run for
30,000 iterations. The trace and autocorrelation plots of the
marginal chains are given in Figures~\ref{fig:tr.acf.x2} and
\ref{fig:tr.acf.y2}. From these plots one may conclude
mixing of the Markov chains, although the large autocorrelations
result in low ESS for the chains. The minimum and maximum mESS
(\ref{eq:mess}) computed using the R package mcmcse for the four
chains are 412 and 469, respectively.

\begin{figure}[h]
\begin{center}
\includegraphics[width=3in]{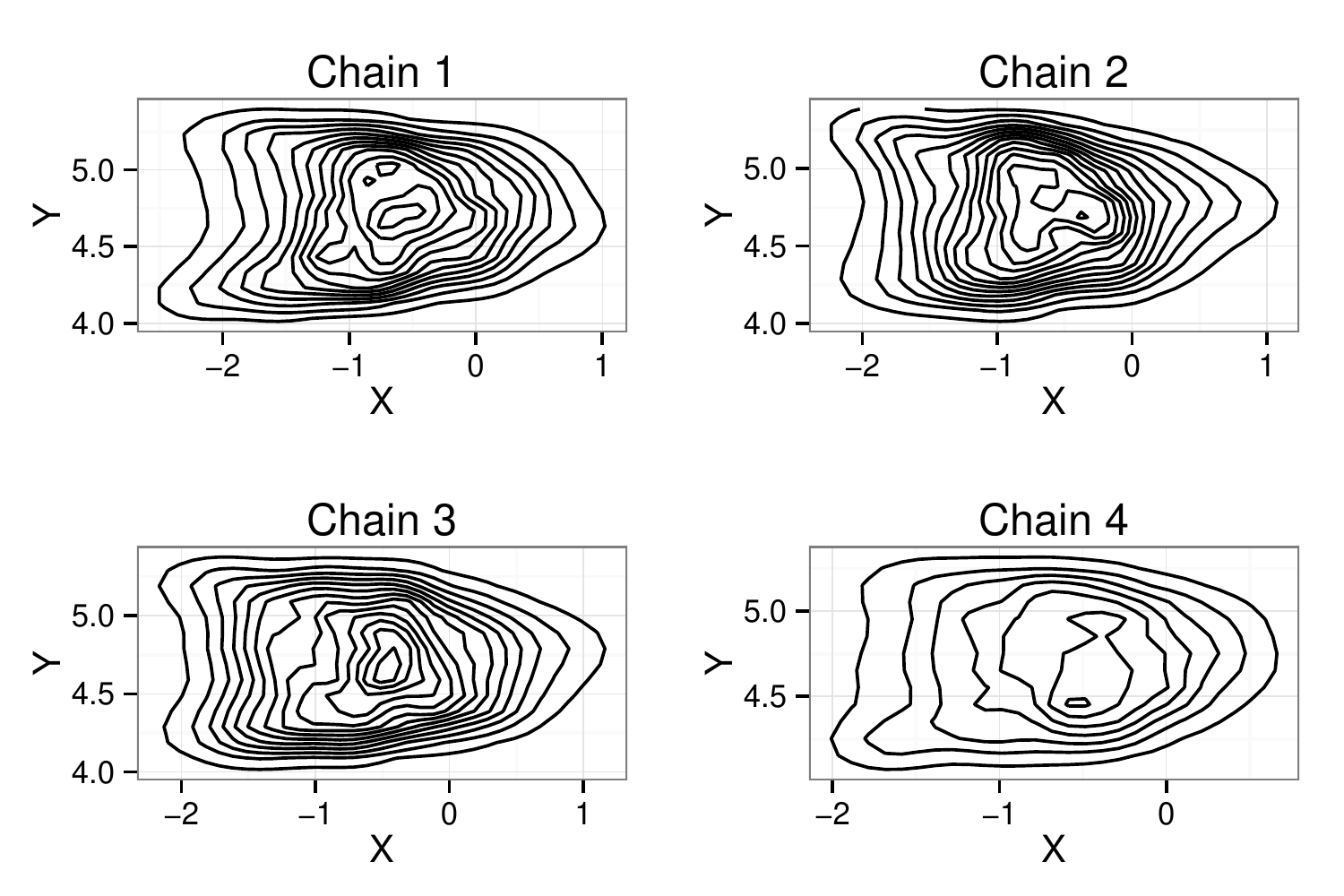}
\end{center}
\caption{Visualizations of the adaptive kernel density estimates of
  the four chains in Case 2 of the sixmodal example. Since the
  bivariate density plots look similar, it fails to provide indication
  of non-convergence of the chains to the stationary distribution.}
\label{fig:c2_chains}
\end{figure}

The adaptive kernel density estimates of the four chains are
visualized in Figure \ref{fig:c2_chains}. This bivariate density plot
does not reveal non-convergence of the chains to the stationary
distribution.
Next, we compute the \cite{gewe:1992}'s and \cite{heid:welc:1983}'s
convergence diagnostics for the identity function $g(x)=x$ for all
four individual chains. At level 0.05, the \cite{gewe:1992} diagnostic
fails to reject the hypothesis of the equality of means from the
beginning and end parts of each chain. Similarly, all chains pass
the \cite{heid:welc:1983} test for stationarity. Thus, both
\cite{gewe:1992}'s and \cite{heid:welc:1983}'s diagnostics fail to
detect the non-convergence of the chains to the target distribution.
Also, the Raftery-Lewis diagnostic fails to distinguish between the chains
in Case 1 and Case 2 as it results in similar burn-in estimates in
both cases.

\begin{figure*}
    \includegraphics[width=0.5\linewidth]{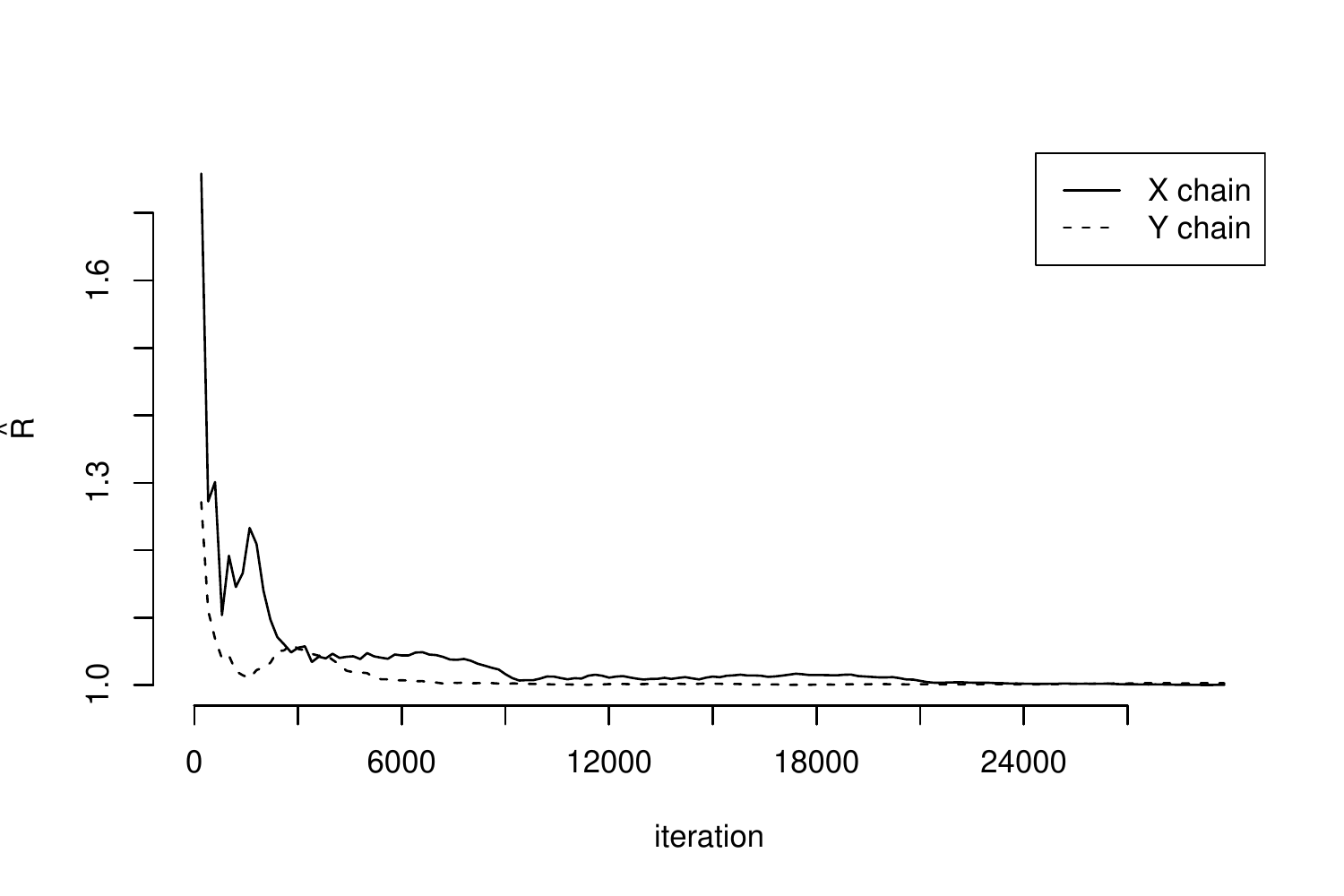}
    \caption{Iterative $\hat{R}$ plot from four parallel chains for
      the sixmodal example in Case 2.}
\label{fig:sixmo:gr.run}
\end{figure*}

We also calculate the PSRF for the marginal chains as well as the
MPSRF for the joint chain based on the four parallel chains as the GR
diagnostic is often used by practitioners for determining burn-in
\citep[][p. 256]{fleg:hara:jone:2008}. The plots of iterative
$\hat{R}$ at increments of 200 iterations are given in
Figure~\ref{fig:sixmo:gr.run}. PSRFs for the marginal chains reach
below 1.1 before 3,000 iterations. The MPSRF (not shown in the plot)
also reaches below 1.1 before 6,000 iterations. Both the PSRF
and MPSRF values are close to one, which is often used as sign of
convergence to stationarity.

Since all four chains are stuck at the same local mode, that is, these
are not run long enough to move between the modes, the convergence
diagnostics, including PSRF, MPSRF get fooled into thinking that the
target distribution is unimodal and hence falsely detect
convergence. \cite{laha:dutt:roy:2017} demonstrate failures of trace
plots, autocorrelation plots and PSRF in diagnosing non-convergence of
MCMC samplers in the context of a statistical model used for analyzing
rank data [See \cite{hobe:roy:robe:2011} for examples of multimodal
targets arising from the popular Bayesian finite mixture models where
empirical convergence diagnostic tools face similar issues.] Since
these diagnostic tools make use of (only) the samples obtained from
the MCMC algorithm, and all observations lie around the same mode,
they fail to diagnose non-convergence. In contrast,
\cite{dixi:roy:2017}'s Tool 2 (\ref{eq:tool2}) uses both MCMC samples
and the target density. Since \cite{dixi:roy:2017}'s Tool 2 requires
only one chain and since the PSRF suggest that the four chains are
similar, we simply choose one of the four chains. Now,
$T_2^\ast = 0.88$ is significantly greater than zero and thus
indicates that the chains are stuck at the same mode.  Furthermore, it
also indicates that 88\% of the target distribution is not yet
captured by the Markov chain. Thus, \cite{dixi:roy:2017}'s Tool 2 is
successful in detecting the divergence of the chains.

\subsection{A Bayesian logistic model}
\label{sec:logi}
In this section, we illustrate MCMC convergence diagnostics in the
context of a real data analysis using a popular statistical model. In
particular, we fit a Bayesian logistic model on the {\it Anguilla
  australis} distribution dataset provided in the R package dismo
\citep{R:dism}. Data are available on a number of sites with presence
or absence of the short-finned eel ({\it Anguilla australis}) in New
Zealand, and some environmental variables at these sites. In
particular, we fit the Anguilla\_train data available in the dismo
package. Here, the response variable is the presence or absence of
short-finned eel, and six other variables are included as
covariates. The six covariates are: summer air temperature (SeqSumT),
distance to coast (DSDist), area with indigenous forest (USNative),
average slope in the upstream catchment (USSlope), maximum downstream
slope (DSMaxSlope) and fishing method (categorical variable with five
classes: electric, mixture, net, spot and trap). Thus the data
set consists of $(y_i, x_i), i=1,\dots,1,000$, where $y_i$ is the $i$th
observation of the response variable taking value 1 (presence) or zero
(absence), and $x_i = (1, \tilde{x}_i)$ is the ten-dimensional
covariate vector, 1 for the intercept and $\tilde{x}_i$ for the other
nine covariates (with four components for the categorical variable
fishing method).  This example was also used by \cite{dixi:roy:2017}
and \cite{boon:merr:krac:2014} to illustrate their MCMC convergence
diagnostic tools.

Denote $\beta =(\beta_0, \beta_1, \dots, \beta_9)$ where $\beta_0$ is
the intercept and $(\beta_1, \dots, \beta_9)$ is the $9 \times 1$ vector of unknown regression
coefficients. We consider the logistic regression model
\[
Y_i|\beta \stackrel{ind}{\sim} \mbox{Bernoulli}(F(x_i^T \beta)), i=1,\dots,1,000,
\]
where $F(\cdot)$ is the cdf of the logistic distribution, that is,
\[
F(x_i^T \beta) = \frac{\exp(x_i^T \beta)}{1+\exp(x_i^T \beta)},i=1,\dots,1,000.
\]
We consider a Bayesian analysis with a diffuse normal prior on $\beta$. Thus,
the posterior density is
\begin{equation}
  \label{eq:logipost}
  \pi(\beta | y) \propto \ell(\beta | y) \phi_{10}(\beta) = \prod_{i=1}^n F(x_i^T \beta)^{y_i} \{1 - F(x_i^T \beta)\}^{1 - y_i} \phi_{10}(\beta),
\end{equation}
where $\ell(\beta | y)$ is the likelihood function and $\phi_{10}(\beta)$ is the density of
$N(\textbf{0}, 100~I_{10})$. The posterior density (\ref{eq:logipost})
is intractable in the sense that means with respect to this density,
which are required for Bayesian inference, are not available in closed
form.

\begin{figure*}
  \begin{minipage}[b]{0.5\linewidth}
    \includegraphics[width=\linewidth]{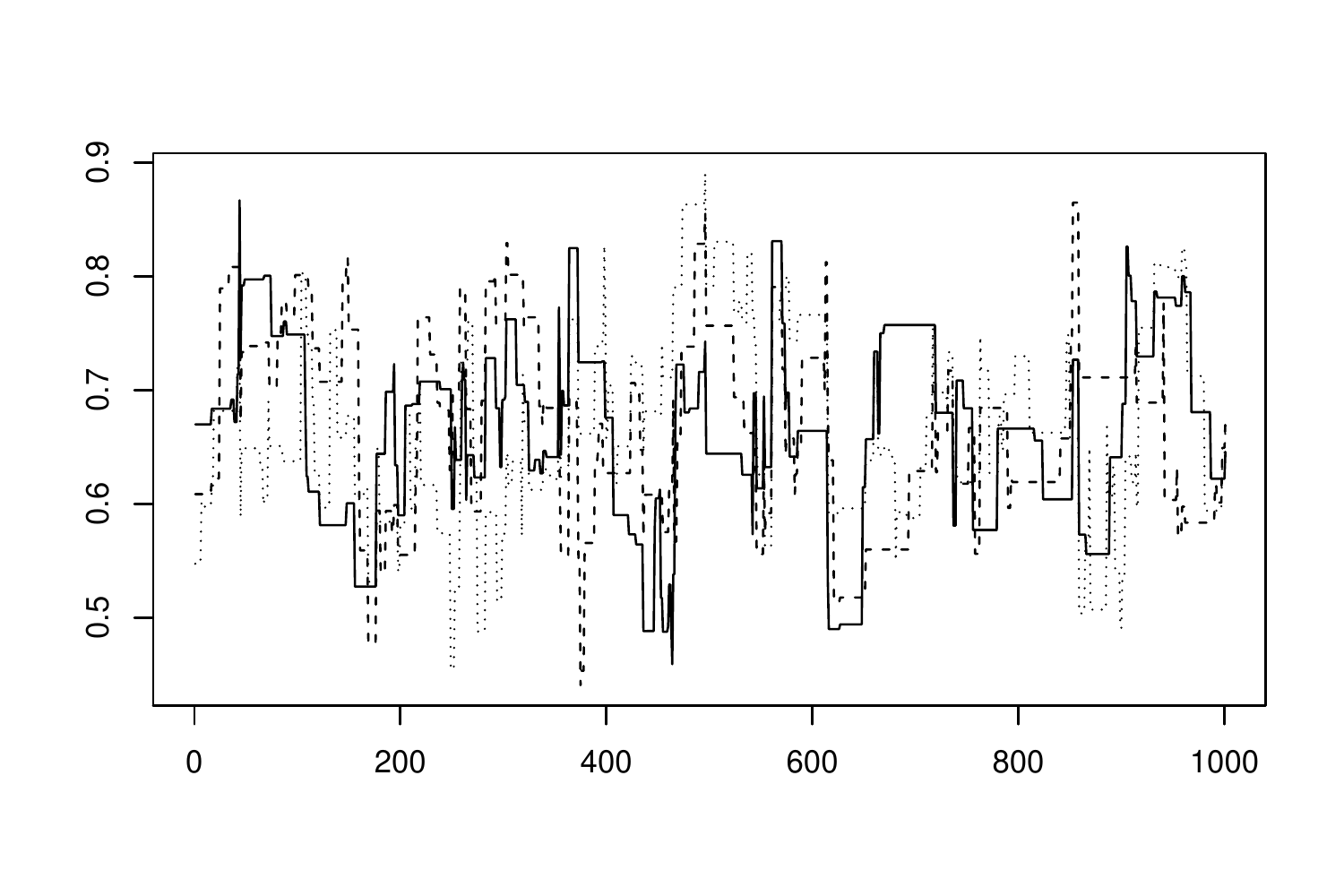}
  \end{minipage}
\begin{minipage}[b]{0.5\linewidth}
    \includegraphics[width=\linewidth]{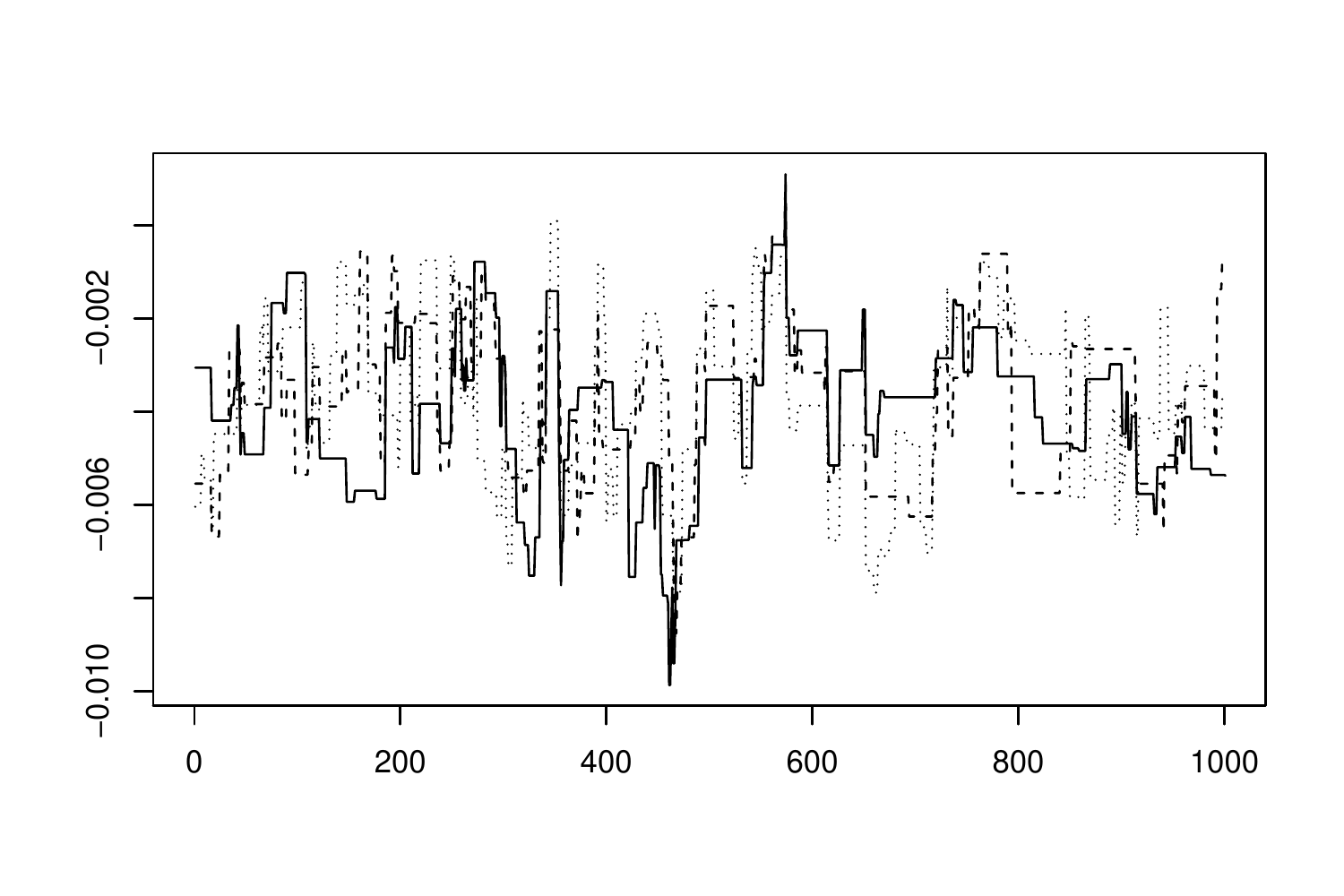}
  \end{minipage}
  \caption{Trace plots of the three chains with default tuning for the
    regression coefficients of summer air temperature (left panel) and
    distance to coast (right panel) for the Bayesian logistic model
    example. The presence of frequent flat bits indicates slow mixing of the Markov chains.}
\label{fig:logit:tracedefault}
\end{figure*}

\begin{figure*}
    \includegraphics[width=\linewidth]{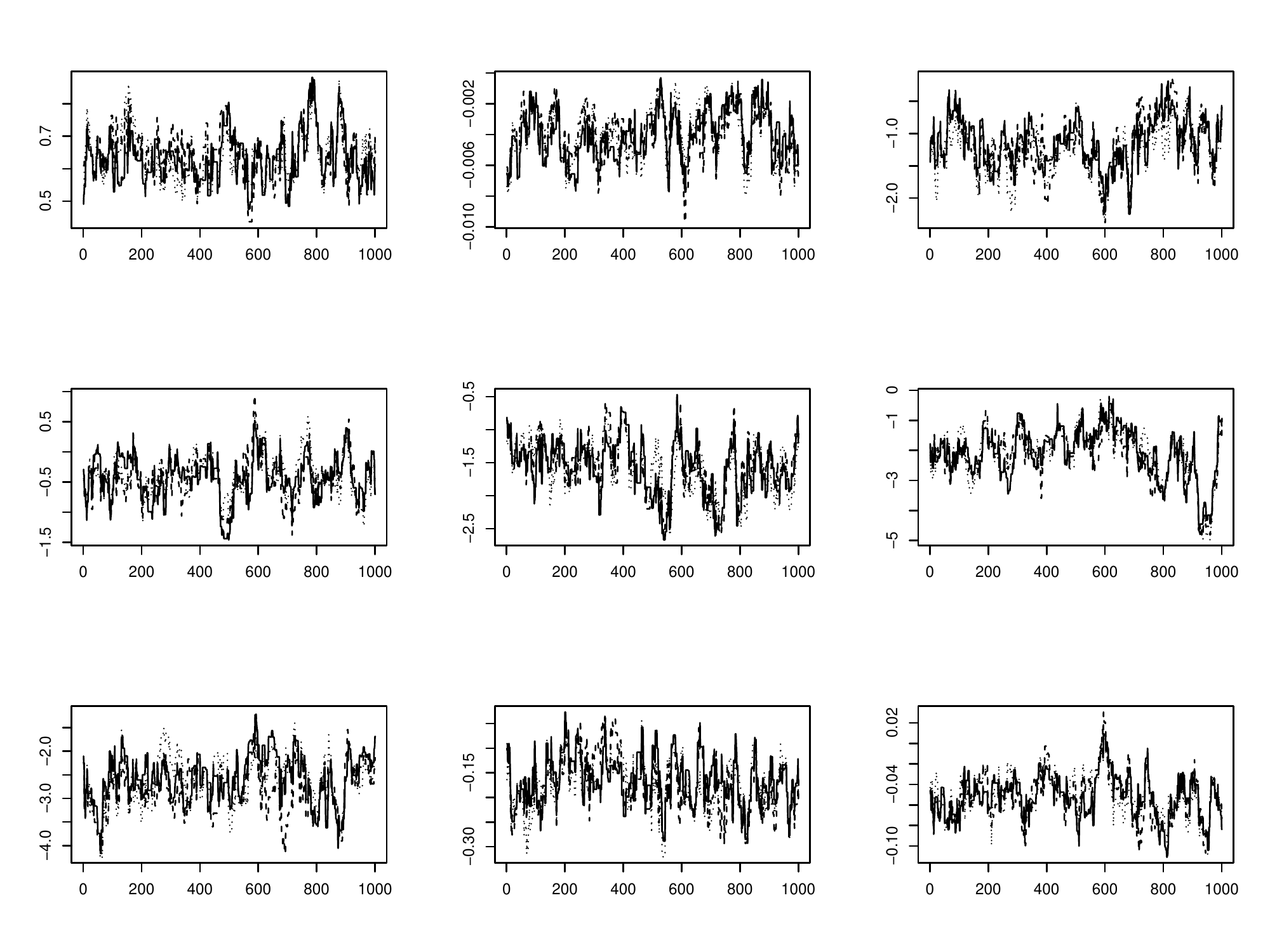}
    \caption{Trace plots of the three chains for the nine regression
      coefficients variables for the Bayesian logistic model
      example. The plots show improved mixing from tuning the
      acceptance rate of the Markov chains.}
\label{fig:logit:trace}
\end{figure*}

\begin{figure*}
    \includegraphics[width=0.5\linewidth]{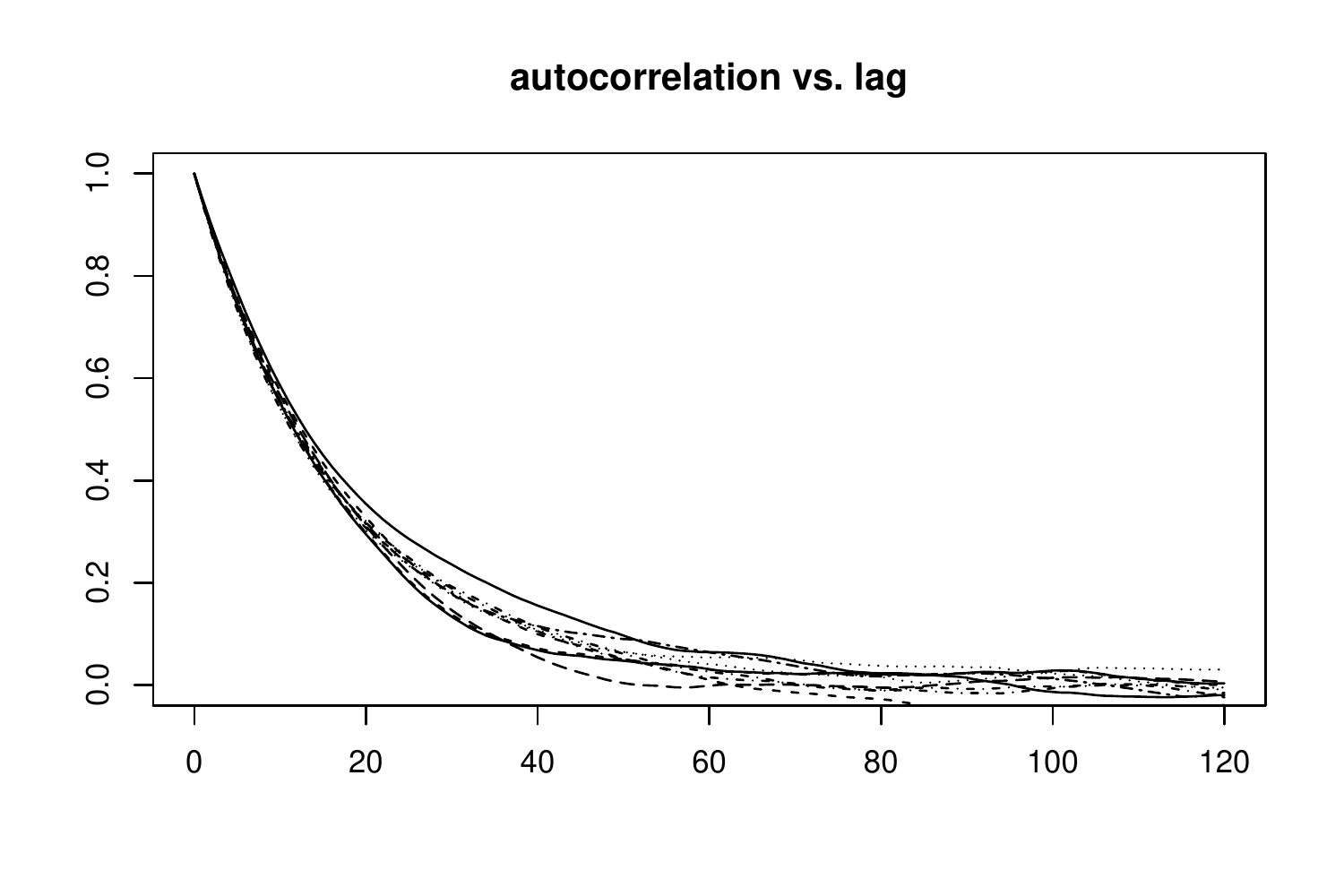}
    \caption{Autocorrelation plots of the ten marginal chains for the Bayesian logistic model example.}
\label{fig:logit:acf}
\end{figure*}

As in \cite{dixi:roy:2017} and \cite{boon:merr:krac:2014}, we use the
MCMClogit function in the R package MCMCpack \citep{R:mcmcp} to draw
MCMC samples from the target density $\pi(\beta | y)$. The maximum
likelihood estimate (MLE) of $\beta$ is the value of the parameter
where the likelihood function $\ell(\beta | y)$ is maximized. Exact
MLE is not available for the logistic likelihood function, neither is
the mode of the posterior density (\ref{eq:logipost}). But, numerical
optimization methods can be used to find an approximate MLE or
posterior mode, which may then be used as starting values. In order to
assess convergence to stationarity, we run three parallel chains with
the default tuning values for 5,000 iterations, one initialized at the
MLE and the other two initialized at points away from the MLE.  Trace
plots of the three chains for the last 1,000 iterations for the
regression coefficients of summer air temperature (left panel) and
distance to coast (right panel) are given in
Figure~\ref{fig:logit:tracedefault}. Trace plots of the other
variables look similar. From these plots we see that, there is not
much overlap between the three parallel chains. From the frequent flat
bits, it follows that the Markov chains move tardily and suffer from
slow mixing. Indeed, the default tuning parameters in the MCMClogit
function result in low (0.11) acceptance rate. We next set the tuning
parameters to achieve around $40\%$ acceptance rate and all analysis
in the remaining section is based on these new tuning values. We run
the three chains longer (30,000 iterations) to obtain reliable ACF
plots. Trace plots of the last 1,000 iterations for each of the three
chains for the nine regression coefficient variables are given in
Figure~\ref{fig:logit:trace}. From the trace plots we see that
convergence of the chains can be further improved. Autocorrelations
for all ten variables for one of the chains based on all 30,000 draws
are given in Figure~\ref{fig:logit:acf}. Autocorrelations for the
other two chains look similar (not included here). Like the trace
plots, the autocorrelation plots also reveal that the Markov chains
suffer from high autocorrelations. It is further corroborated by the
mESS values, which are less than 1,000 for all the three
chains. To sample from (\ref{eq:logipost}) one may use an alternative
MCMC sampler, e.g., the P\'olya-Gamma Gibbs sampler
\citep{pols:scot:wind:2013}, which is known to be geometrically
ergodic \citep{wang:roy:2018b, choi:hobe:2013}. Here we do not use the
P\'olya-Gamma Gibbs sampler as our goal is to illustrate the
convergence diagnostic methods. The MPSRF reaches close to one before
30,000 iterations. Since the Markov chains are 10-dimensional, to
maintain an overall type 1 error rate of $\alpha=0.05$, using
Bonferroni's correction, \cite{dixi:roy:2017} advocate the cutoff
point $0.01$ for the KL Tool 1 for marginal chains. For each of the
ten variables, the maximum symmetric KL divergence among the three
pairs of chains is computed. It turns out that the marginal chains do
not pass the KL Tool 1 test as the maximum symmetric KL divergence
takes the value 7.26 for the variable USSlope. After 30,000
iterations, all marginal chains pass the \cite{heid:welc:1983}
stationarity test. On the other hand, for each of the three parallel
chains, for some of the variables, the \cite{gewe:1992} $Z$ test turns
out to be significant at $0.05$ level. Next, we run the chains for
another 40,000 iterations. For the last 40,000 iterations, all
marginal chains pass the \cite{gewe:1992} $Z$ test, as well as the
KL Tool 1 test. Also, based on these 40,000 iterations, the maximum
burn-in estimate from the  Raftery-Lewis diagnostic (with
$\epsilon=0.005, \alpha=0.05$) over different quantiles
$(q=0.1,\dots,0.9)$ is less than 100 for all 10 variables.  We thus
use $n' =$70,000 as the burn-in value.

After removing the first 70,000 iterations as initial burn-in, each of
the three chains is run for an additional 15,000 iterations. Table
\ref{tab:logit_tab} presents the PSRF and the maximum symmetric KL
divergence [\cite{dixi:roy:2017}'s KL Tool 1] values based on three
parallel chains for all 10 variables. The half-widths of the 95\%
confidence intervals based on the first chain (started at the MLE) are
also tabulated in Table \ref{tab:logit_tab}. All values are given up
to three decimal places. MPSRF takes the value 1.004.  For the three
chains mESS takes values 515, 520 and 502, respectively.
High cross-correlation between the Intercept and SeqSumT regression
coefficient parameters (-0.984) and between USNative and USSlope
(-0.558) suggest that mixing of the Markov chain can improve if it is run on
an appropriate lower dimensional space (that is, after dropping some
variables) or a reparameterization is used. From Table
\ref{tab:logit_tab}, we see that all marginal chains pass the KL Tool
1 diagnostic.  Also, all PSRF values as well as the MPSRF value reach
below the cutoff 1.1.  On the other hand, the maximum half-width
among the 10 regression parameters is $0.112$, much larger than the
cutoff 0.01. Doing a simple sample size calculation, based on the
pilot sample size 15,000, we find that we need
15,000$\times(0.112/0.01)^2=$ 1,881,600 samples for obtaining
confidence intervals with half-widths below 0.01.

\begin{table}[h]
\caption{Application of various MCMC convergence diagnostic tools to the Bayesian logistic model.}
\centering
\begin{tabular}{|l|lll|lll|}
\hline
\textbf{Variable} &$\mathbf{\hat{R}}$&\textbf{half-width}&\textbf{Tool 1}\\
\hline
Intercept  & 1.000 & 0.112 & 0.008 \\
SeqSumT  & 1.000 & 0.006 & 0.007  \\
 DSDist & 1.001 & 0.000 & 0.005\\
 USNative & 1.001 & 0.025 & 0.004    \\
 M - mix & 1.000 & 0.031 & 0.005    \\
 M - net & 1.001 & 0.031 & 0.004 \\
 M - spot & 1.000 & 0.048 & 0.004   \\
 M - trap & 1.002 & 0.051 & 0.004 \\
 DSMaxSlope & 1.000 & 0.005 & 0.007 \\
 USSlope & 1.001 & 0.002 & 0.004 \\
\hline
\end{tabular}
\label{tab:logit_tab}
\end{table}

Finally, we run one of the chains (the chain started at the MLE) for
1,881,600 iterations after a burn-in of $n'=$70,000 iterations. Thus
the chain is stopped after $n^*=$1,951,600 iterations. In this case,
as expected, the maximum half-width of the $95\%$ confidence interval
is below $0.01$. An estimate of mESS calculated using the mcmcse
package is 55,775 which is larger than the cutoff value 55,191 given
in (\ref{eq:messcut}) for $p=10$, $\alpha =0.05$ and
$\varepsilon =0.02$. On the other hand, the chain needs to be run
longer to achieve the cutoff value 220,766 (\ref{eq:messcut})
corresponding to $\varepsilon =0.01$.  Table~\ref{tab:logit_est} gives
the estimates of posterior means of all regression coefficients
and their corresponding Monte Carlo standard errors (SE).

\begin{table}[h]
\caption{Estimates of posterior means and standard errors of regression coefficients for the Bayesian logistic model.}
\centering
\begin{tabular}{|l|llllllllll|}
\hline
\textbf{Variable} &
$\beta_0$ &
$\beta_1$ &
$\beta_2$&
$\beta_3$&
$\beta_4$&
$\beta_5$&
$\beta_6$&
$\beta_7$&
 $\beta_8$&
 $\beta_9$\\
\hline
\textbf{Estimate} &
-10.46 &
  0.66&
 -0.00&
 -1.17&
 -0.47&
 -1.53&
 -1.83&
 -2.59&
 -0.17&
 -0.05\\
\hline
\textbf{SE $\times 10^3$} &
5.73&
0.32&
0.00&
1.52&
1.46&
1.65&
2.76&
2.35&
0.24&
0.08\\
\hline
\end{tabular}
\label{tab:logit_est}
\end{table}

\newpage
\section{Conclusions and discussion}
In this article, we discuss several measures for diagnosing
convergence of Monte Carlo Markov chains to stationarity as well as
convergence of the sample averages based on these chains. Detection of
the first is often used to decide a suitable burn-in period, while the
second leads to termination of the MCMC simulation. Analytical upper
bounds to the TV norm required to obtain an honest burn-in maybe
difficult to find in practice or may lead to very conservative burn-in
values. On the other hand, empirical diagnostics can falsely detect
convergence when the chains are not run long enough to move between
the modes. For the chains initialized at high density density regions,
there is no need for burn-in.  If the global mode of the target
density can be (approximately) found by optimization then it can be
used as the starting value.

Some of the empirical diagnostics for convergence of sample averages may
prematurely terminate the simulation and the resulting inference can
be far from the truth. Thus, use of fixed-width and ESS---based
stopping rules is recommended. Most of the quantitative convergence
diagnostics assumes a Markov chain CLT. While demonstrating the
existence of a Markov chain CLT requires some rigorous theoretical
analysis of the Markov chain, given the great amount of work done in
this direction, validating honest stopping rules does not present as
much of an obstacle as in the past.

None of the three examples discussed here use thinning. Thinning,
that is, discarding all but every $k$th observation, is often used by
MCMC practitioners to reduce high autocorrelations present in the
Markov chain samples. Since it wastes too many samples, it should be
used only if computer storage of the samples is an issue or evaluating
the functions of interest ($g$) is more expensive than sampling the
Markov chain. If thinning is used, convergence diagnostics can be used
on the thinned samples.

Some convergence diagnostic tools use parallel chains initialized at
different points, or two parts of a single chain. In the presence of
multiple modes, if the initial points of the parallel chains are not
in distinct high-density regions, or the chain is not run long enough
to move between the modes, the diagnostics fail to detect the
non-convergence. Thus, single long runs should be used to make final
inference. Running the chain longer may also result in discovering new
parts of the support of the target distribution. In contrast,
recently, \cite{jaco:olea:atch:2017} propose a method for
parallelizing MCMC computations using couplings of Markov chains.

Practitioners should be careful while depending purely on empirical
convergence diagnostic tools, especially if the presence of multiple
modes is suspected. Empirical diagnostics cannot detect convergence
with certainty. Also, if the target is incorrectly assumed to be a
proper density, the empirical diagnostic tools may not provide a red
flag indicating its impropriety \citep{athr:roy:2014b,
  hobe:case:1996}. Over the past two decades, much research has been
done to provide honest Monte Carlo sample size calculation for myriad
MCMC algorithms for common statistical models. However, theoretical
analysis of MCMC algorithms is an ongoing area of research and further
important work needs to be done. A potential future study involves
theoretically verifying the convergence (to zero) of
\cite{dixi:roy:2017}'s statistics based on the KL divergence. Another
possible research problem is to construct theoretically valid and
computationally efficient MCMC convergence diagnostics in
ultrahigh-dimensional settings.

\section*{ACKNOWLEDGMENTS}

The author thanks one anonymous editor for careful and detailed
comments on an earlier version of the manuscript. The author thanks
Evangelos Evangelou, Mark Kaiser and Dootika Vats for helpful
comments. These valuable suggestions have substantially improved the
article. The author also thanks Chris Oats for suggesting two references
and Anand Dixit for providing some R codes used in the second and
third examples.


\bibliographystyle{ims}
\bibliography{ref}

\begin{thebibliography}{74}
\expandafter\ifx\csname natexlab\endcsname\relax\def\natexlab#1{#1}\fi
\expandafter\ifx\csname url\endcsname\relax
  \def\url#1{\texttt{#1}}\fi
\expandafter\ifx\csname urlprefix\endcsname\relax\def\urlprefix{URL }\fi
\providecommand{\eprint}[2][]{\url{#2}}

\bibitem[{Andrieu et~al.(2015)Andrieu, Fort and Vihola}]{andr:fort:viho:2015}
\textsc{Andrieu, C.}, \textsc{Fort, G.} and \textsc{Vihola, M.} (2015).
\newblock Quantitative convergence rates for subgeometric {M}arkov chains.
\newblock \textit{Journal of Applied Probability}, \textbf{52} 391--404.

\bibitem[{Asmussen and Glynn(2011)}]{asmu:glyn:2011}
\textsc{Asmussen, S.} and \textsc{Glynn, P.~W.} (2011).
\newblock A new proof of convergence of {MCMC} via the ergodic theorem.
\newblock \textit{Statistics and Probability Letters}, \textbf{81} 1482--1485.

\bibitem[{Athreya and Roy(2014)}]{athr:roy:2014b}
\textsc{Athreya, K.~B.} and \textsc{Roy, V.} (2014).
\newblock Monte {C}arlo methods for improper target distributions.
\newblock \textit{Electronic Journal of Statistics}, \textbf{8} 2664--2692.

\bibitem[{Baxendale(2005)}]{baxe:2005}
\textsc{Baxendale, P.~H.} (2005).
\newblock Renewal theory and computable convergence rates for geometrically
  ergodic {M}arkov chains.
\newblock \textit{The Annals of Applied Probability}, \textbf{15} 700--738.

\bibitem[{Boone et~al.(2014)Boone, Merrick and Krachey}]{boon:merr:krac:2014}
\textsc{Boone, E.}, \textsc{Merrick, J.} and \textsc{Krachey, M.} (2014).
\newblock A {H}ellinger distance approach to {MCMC} diagnostics.
\newblock \textit{Journal of Statistical Computation and Simulation},
  \textbf{84} 833--849.

\bibitem[{Brooks and Gelman(1998)}]{broo:gelm:1998}
\textsc{Brooks, S.~P.} and \textsc{Gelman, A.} (1998).
\newblock General methods for monitoring convergence of iterative simulations.
\newblock \textit{Journal of Computational and Graphical Statistics},
  \textbf{7} 434--455.

\bibitem[{Brooks and Roberts(1998)}]{broo:robe:1998}
\textsc{Brooks, S.~P.} and \textsc{Roberts, G.~O.} (1998).
\newblock Assessing convergence of {M}arkov chain {M}onte {C}arlo algorithms.
\newblock \textit{Statistics and Computing}, \textbf{8} 319--335.

\bibitem[{Chakraborty and Khare(2017)}]{chak:khar:2017}
\textsc{Chakraborty, S.} and \textsc{Khare, K.} (2017).
\newblock Convergence properties of {G}ibbs samplers for {B}ayesian probit
  regression with proper priors.
\newblock \textit{Electronic Journal of Statistics}, \textbf{11} 177--210.

\bibitem[{Choi and Hobert(2013)}]{choi:hobe:2013}
\textsc{Choi, H.~M.} and \textsc{Hobert, J.~P.} (2013).
\newblock The {P}olya-{G}amma {G}ibbs sampler for {B}ayesian logistic
  regression is uniformly ergodic.
\newblock \textit{Electronic Journal of Statistics}, \textbf{7} 2054--2064.

\bibitem[{Christensen(2004)}]{chri:2004}
\textsc{Christensen, O.~F.} (2004).
\newblock Monte {C}arlo maximum likelihood in model based geostatistics.
\newblock \textit{Journal of Computational and Graphical Statistics},
  \textbf{13} 702--718.

\bibitem[{Cowles and Carlin(1996)}]{cowl:carl:1996}
\textsc{Cowles, M.~K.} and \textsc{Carlin, B.~P.} (1996).
\newblock Markov chain {M}onte {C}arlo convergence diagnostics: a comparative
  review.
\newblock \textit{Journal of the American Statistical Association}, \textbf{91}
  883--904.

\bibitem[{Dixit and Roy(2017)}]{dixi:roy:2017}
\textsc{Dixit, A.} and \textsc{Roy, V.} (2017).
\newblock M{C}{M}{C} diagnostics for higher dimensions using {K}ullback
  {L}eibler divergence.
\newblock \textit{Journal of Statistical Computation and Simulation},
  \textbf{87} 2622--2638.

\bibitem[{Doss et~al.(2014)Doss, Flegal, Jones and
  Neath}]{doss:fleg:jone:neat:2014}
\textsc{Doss, C.~R.}, \textsc{Flegal, J.~M.}, \textsc{Jones, G.~L.} and
  \textsc{Neath, R.~C.} (2014).
\newblock Markov chain {M}onte {C}arlo estimation of quantiles.
\newblock \textit{Electronic Journal of Statistics}, \textbf{8} 2448--2478.

\bibitem[{Durmus and Moulines(2015)}]{durm:moul:2015}
\textsc{Durmus, A.} and \textsc{Moulines, {\'E}.} (2015).
\newblock Quantitative bounds of convergence for geometrically ergodic {M}arkov
  chain in the {W}asserstein distance with application to the {M}etropolis
  {A}djusted {L}angevin {A}lgorithm.
\newblock \textit{Statistics and Computing}, \textbf{25} 5--19.

\bibitem[{Evangelou and Roy(2019)}]{r:geoBayes}
\textsc{Evangelou, E.} and \textsc{Roy, V.} (2019).
\newblock \textit{geoBayes: Analysis of Geostatistical Data using Bayes and
  Empirical Bayes Methods}.
\newblock R package version 0.6.2,
  \urlprefix\url{https://CRAN.R-project.org/package=geoBayes}.

\bibitem[{Flegal and Gong(2015)}]{fleg:gong:2015}
\textsc{Flegal, J.~M.} and \textsc{Gong, L.} (2015).
\newblock Relative fixed-width stopping rules for {M}arkov chain {M}onte
  {C}arlo simulations.
\newblock \textit{Statistica Sinica} 655--675.

\bibitem[{Flegal et~al.(2008)Flegal, Haran and Jones}]{fleg:hara:jone:2008}
\textsc{Flegal, J.~M.}, \textsc{Haran, M.} and \textsc{Jones, G.~L.} (2008).
\newblock Markov chain {M}onte {C}arlo: {C}an we trust the third significant
  figure?
\newblock \textit{Statistical Science}, \textbf{23} 250--260.

\bibitem[{Flegal et~al.(2012)Flegal, Hughes, Vats and Dai}]{R:mcmcse}
\textsc{Flegal, J.~M.}, \textsc{Hughes, J.}, \textsc{Vats, D.} and \textsc{Dai,
  N.} (2012).
\newblock \textit{mcmcse: Monte Carlo standard errors for MCMC}.
\newblock R package version 0.1,
  \urlprefix\url{http://CRAN.R-project.org/package=mcmcse}.

\bibitem[{Flegal and Jones(2010)}]{fleg:jone:2010}
\textsc{Flegal, J.~M.} and \textsc{Jones, G.~L.} (2010).
\newblock Batch means and spectral variance estimators in {M}arkov chain
  {M}onte {C}arlo.
\newblock \textit{The Annals of Statistics}, \textbf{38} 1034--1070.

\bibitem[{Fort et~al.(2003)Fort, Moulines, Roberts and
  Rosenthal}]{fort:moul:robe:rose:2003}
\textsc{Fort, G.}, \textsc{Moulines, E.}, \textsc{Roberts, G.} and
  \textsc{Rosenthal, J.} (2003).
\newblock On the geometric ergodicity of hybrid samplers.
\newblock \textit{Journal of Applied Probability}, \textbf{40} 123--146.

\bibitem[{Gelman et~al.(2014)Gelman, Carlin, Stern, Dunson, Vehtari and
  Rubin}]{gelm:carl:ster:duns:veht:rubi:2014}
\textsc{Gelman, A.}, \textsc{Carlin, J.~B.}, \textsc{Stern, H.~S.},
  \textsc{Dunson, D.~B.}, \textsc{Vehtari, A.} and \textsc{Rubin, D.~B.}
  (2014).
\newblock \textit{Bayesian data analysis}.
\newblock Chapman and Hall/CRC.

\bibitem[{Gelman and Rubin(1992)}]{gelm:rubi:1992}
\textsc{Gelman, A.} and \textsc{Rubin, D.~B.} (1992).
\newblock Inference from iterative simulation using multiple sequences.
\newblock \textit{Statistical Science}, \textbf{7} 457--472.

\bibitem[{Geweke(1992)}]{gewe:1992}
\textsc{Geweke, J.} (1992).
\newblock Evaluating the accuracy of sampling-based approaches to calculating
  posterior moments.
\newblock In \textit{Bayesian {S}tatistics 4} (J.~M. Bernado, J.~O. Berger,
  A.~P. Dawid and A.~F.~M. Smith, eds.). Clarendon Press, Oxford, UK, 169--193.

\bibitem[{Geyer(2011)}]{geye:2011}
\textsc{Geyer, C.~J.} (2011).
\newblock \textit{Handbook of {M}arkov chain {M}onte {C}arlo}, chap.
  Introduction to {M}arkov chain {M}onte {C}arlo.
\newblock CRC Press, Boca Raton, FL, 3--48.

\bibitem[{Geyer and Johnson(2017)}]{r:mcmc}
\textsc{Geyer, C.~J.} and \textsc{Johnson, L.~T.} (2017).
\newblock \textit{mcmc: Markov Chain Monte Carlo}.
\newblock R package version 0.9-5,
  \urlprefix\url{https://CRAN.R-project.org/package=mcmc}.

\bibitem[{Geyer and Thompson(1995)}]{geye:thomp:1995}
\textsc{Geyer, C.~J.} and \textsc{Thompson, E.~A.} (1995).
\newblock Annealing {M}arkov chain {M}onte {C}arlo with applications to
  ancestral inference.
\newblock \textit{Journal of the American Statistical Association}, \textbf{90}
  909--920.

\bibitem[{Glynn and Whitt(1992)}]{glyn:whit:1992}
\textsc{Glynn, P.~W.} and \textsc{Whitt, W.} (1992).
\newblock The asymptotic validity of sequential stopping rules for stochastic
  simulations.
\newblock \textit{The Annals of Applied Probability}, \textbf{2} 180--198.

\bibitem[{Gong and Flegal(2016)}]{gong:fleg:2016}
\textsc{Gong, L.} and \textsc{Flegal, J.~M.} (2016).
\newblock A practical sequential stopping rule for high-dimensional {M}arkov
  chain {M}onte {C}arlo.
\newblock \textit{Journal of Computational and Graphical Statistics},
  \textbf{25} 684--700.

\bibitem[{Gorham and Mackey(2015)}]{gorh:mack:2015}
\textsc{Gorham, J.} and \textsc{Mackey, L.} (2015).
\newblock Measuring sample quality with {S}tein's method.
\newblock In \textit{Advances in Neural Information Processing Systems}.
  226--234.

\bibitem[{Hadfield(2010)}]{r:MCMCglmm}
\textsc{Hadfield, J.~D.} (2010).
\newblock {MCMC} methods for multi-response generalized linear mixed models:
  The {MCMCglmm} {R} package.
\newblock \textit{Journal of Statistical Software}, \textbf{33} 1--22.
\newblock \urlprefix\url{http://www.jstatsoft.org/v33/i02/}.

\bibitem[{Heidelberger and Welch(1983)}]{heid:welc:1983}
\textsc{Heidelberger, P.} and \textsc{Welch, P.~D.} (1983).
\newblock Simulation run length control in the presence of an initial
  transient.
\newblock \textit{Operations Research}, \textbf{31} 1109--1144.

\bibitem[{Hijmans et~al.(2016)Hijmans, Phillips, Leathwick and Elith}]{R:dism}
\textsc{Hijmans, R.~J.}, \textsc{Phillips, S.}, \textsc{Leathwick, J.} and
  \textsc{Elith, J.} (2016).
\newblock \textit{dismo: Species Distribution Modeling}.
\newblock R package version 1.0-15.

\bibitem[{Hobert and Casella(1996)}]{hobe:case:1996}
\textsc{Hobert, J.~P.} and \textsc{Casella, G.} (1996).
\newblock The effect of improper priors on {G}ibbs sampling in hierarchical
  linear mixed models.
\newblock \textit{Journal of the American Statistical Association}, \textbf{91}
  1461--1473.

\bibitem[{Hobert et~al.(2002)Hobert, Jones, Presnell and
  Rosenthal}]{hobe:jone:pres:rose:2002}
\textsc{Hobert, J.~P.}, \textsc{Jones, G.~L.}, \textsc{Presnell, B.} and
  \textsc{Rosenthal, J.~S.} (2002).
\newblock On the applicability of regenerative simulation in {M}arkov chain
  {M}onte {C}arlo.
\newblock \textit{Biometrika}, \textbf{89} 731--743.

\bibitem[{Hobert et~al.(2018)Hobert, Jung, Khare and
  Qin}]{hobe:jung:khar:qin:2018}
\textsc{Hobert, J.~P.}, \textsc{Jung, Y.~J.}, \textsc{Khare, K.} and
  \textsc{Qin, Q.} (2018).
\newblock Convergence analysis of {MCMC} algorithms for {B}ayesian multivariate
  linear regression with non-{G}aussian errors.
\newblock \textit{Scandinavian Journal of Statistics}, \textbf{45} 513--533.

\bibitem[{Hobert et~al.(2011)Hobert, Roy and Robert}]{hobe:roy:robe:2011}
\textsc{Hobert, J.~P.}, \textsc{Roy, V.} and \textsc{Robert, C.~P.} (2011).
\newblock Improving the convergence properties of the data augmentation
  algorithm with an application to {B}ayesian mixture modelling.
\newblock \textit{Statistical Science}, \textbf{26} 332--351.

\bibitem[{Jacob et~al.(2017)Jacob, O'Leary and
  Atchad{\'e}}]{jaco:olea:atch:2017}
\textsc{Jacob, P.~E.}, \textsc{O'Leary, J.} and \textsc{Atchad{\'e}, Y.~F.}
  (2017).
\newblock Unbiased {M}arkov chain {M}onte {C}arlo with couplings.
\newblock \textit{arXiv preprint arXiv:1708.03625}.

\bibitem[{Jones(2004)}]{jone:2004}
\textsc{Jones, G.~L.} (2004).
\newblock On the {M}arkov chain central limit theorem.
\newblock \textit{Probability Surveys}, \textbf{1} 299--320.

\bibitem[{Jones et~al.(2006)Jones, Haran, Caffo and
  Neath}]{jone:hara:caff:neat:2006}
\textsc{Jones, G.~L.}, \textsc{Haran, M.}, \textsc{Caffo, B.~S.} and
  \textsc{Neath, R.} (2006).
\newblock Fixed-width output analysis for {M}arkov chain {M}onte {C}arlo.
\newblock \textit{Journal of the American Statistical Association},
  \textbf{101} 1537--1547.

\bibitem[{Jones and Hobert(2001)}]{jone:hobe:2001}
\textsc{Jones, G.~L.} and \textsc{Hobert, J.~P.} (2001).
\newblock Honest exploration of intractable probability distributions via
  {M}arkov chain {M}onte {C}arlo.
\newblock \textit{Statistical Science}, \textbf{16} 312--34.

\bibitem[{Jones and Hobert(2004)}]{jone:hobe:2004}
\textsc{Jones, G.~L.} and \textsc{Hobert, J.~P.} (2004).
\newblock Sufficient burn-in for {G}ibbs samplers for a hierarchical random
  effects model.
\newblock \textit{The Annals of Statistics}, \textbf{32} 784--817.

\bibitem[{Khare and Hobert(2012)}]{khar:hobe:2012}
\textsc{Khare, K.} and \textsc{Hobert, J.~P.} (2012).
\newblock Geometric ergodicity of the {G}ibbs sampler for {B}ayesian quantile
  regression.
\newblock \textit{Journal of Multivariate Analysis}, \textbf{112} 108--116.

\bibitem[{Khare and Hobert(2013)}]{khar:hobe:2013}
\textsc{Khare, K.} and \textsc{Hobert, J.~P.} (2013).
\newblock Geometric ergodicity of {B}ayesian lasso.
\newblock \textit{Electronic Journal of Statistics}, \textbf{7} 2150--2163.

\bibitem[{Laha et~al.(2016)Laha, Dutta and Roy}]{laha:dutt:roy:2017}
\textsc{Laha, A.}, \textsc{Dutta, S.} and \textsc{Roy, V.} (2016).
\newblock A novel sandwich algorithm for empirical {B}ayes analysis of rank
  data.
\newblock \textit{Statistics and its Interface}, \textbf{10} 543--556.

\bibitem[{Leman et~al.(2009)Leman, Chen and Lavine}]{lem:chen:lavi:2009}
\textsc{Leman, S.~C.}, \textsc{Chen, Y.} and \textsc{Lavine, M.} (2009).
\newblock The multiset sampler.
\newblock \textit{Journal of the American Statistical Association},
  \textbf{104} 1029--1041.

\bibitem[{Martin et~al.(2011)Martin, Quinn and Park}]{R:mcmcp}
\textsc{Martin, A.~D.}, \textsc{Quinn, K.~M.} and \textsc{Park, J.~H.} (2011).
\newblock {MCMCpack}: {M}arkov chain {M}onte {C}arlo in {R}.
\newblock \textit{Journal of Statistical Software}, \textbf{42} 22.

\bibitem[{Mengersen and Tweedie(1996)}]{meng:twee:1996}
\textsc{Mengersen, K.} and \textsc{Tweedie, R.~L.} (1996).
\newblock Rates of convergence of the {H}astings and {M}etropolis algorithms.
\newblock \textit{The Annals of Statistics}, \textbf{24} 101--121.

\bibitem[{Mengersen et~al.(1999)Mengersen, Robert and
  Guihenneuc-Jouyaux}]{meng:robe:guih:1999}
\textsc{Mengersen, K.~L.}, \textsc{Robert, C.~P.} and
  \textsc{Guihenneuc-Jouyaux, C.} (1999).
\newblock {MCMC} convergence diagnostics: a reviewww.
\newblock \textit{Bayesian statistics}, \textbf{6} 415--440.

\bibitem[{Meyn and Tweedie(1993)}]{meyn:twee:1993}
\textsc{Meyn, S.~P.} and \textsc{Tweedie, R.~L.} (1993).
\newblock \textit{Markov {C}hains and {S}tochastic {S}tability}.
\newblock Springer Verlag, London.

\bibitem[{Mykland et~al.(1995)Mykland, Tierney and Yu}]{mykl:tier:yu:1995}
\textsc{Mykland, P.}, \textsc{Tierney, L.} and \textsc{Yu, B.} (1995).
\newblock Regeneration in {M}arkov chain samplers.
\newblock \textit{Journal of the American Statistical Association}, \textbf{90}
  233--41.

\bibitem[{Peltonen et~al.(2009)Peltonen, Venna and Kaski}]{pelt:venn:kask:2009}
\textsc{Peltonen, J.}, \textsc{Venna, J.} and \textsc{Kaski, S.} (2009).
\newblock Visualizations for assessing convergence and mixing of {M}arkov chain
  {M}onte {C}arlo simulations.
\newblock \textit{Computational Statistics and Data Analysis}, \textbf{53}
  4453--4470.

\bibitem[{Plummer et~al.(2006)Plummer, Best, Cowles and Vines}]{R:coda}
\textsc{Plummer, M.}, \textsc{Best, N.}, \textsc{Cowles, K.} and \textsc{Vines,
  K.} (2006).
\newblock Coda: convergence diagnosis and output analysis for {MCMC}.
\newblock \textit{R news}, \textbf{6} 7--11.

\bibitem[{Polson et~al.(2013)Polson, Scott and Windle}]{pols:scot:wind:2013}
\textsc{Polson, N.~G.}, \textsc{Scott, J.~G.} and \textsc{Windle, J.} (2013).
\newblock Bayesian inference for logistic models using {P}{\'o}lya-{G}amma
  latent variables.
\newblock \textit{Journal of the American statistical Association},
  \textbf{108} 1339--1349.

\bibitem[{Qin and Hobert(2019)}]{qin:hobe:2019}
\textsc{Qin, Q.} and \textsc{Hobert, J.~P.} (2019).
\newblock Geometric convergence bounds for {M}arkov chains in {W}asserstein
  distance based on generalized drift and contraction conditions.
\newblock \textit{arXiv preprint arXiv:1902.02964}.

\bibitem[{{R Core Team}(2018)}]{r}
\textsc{{R Core Team}} (2018).
\newblock \textit{R: A Language and Environment for Statistical Computing}.
\newblock R Foundation for Statistical Computing, Vienna, Austria.
\newblock \urlprefix\url{https://www.R-project.org/}.

\bibitem[{Raftery and Lewis(1992)}]{raft:lewi:1992}
\textsc{Raftery, A.~E.} and \textsc{Lewis, S.~M.} (1992).
\newblock How many iterations in the {G}ibbs sampler?
\newblock In \textit{Bayesian {S}tatistics 4} (J.~M. Bernado, J.~O. Berger,
  A.~P. Dawid and A.~F.~M. Smith, eds.). Clarendon Press, Oxford, UK, 763--773.

\bibitem[{Robert and Casella(2004)}]{robe:case:2004}
\textsc{Robert, C.} and \textsc{Casella, G.} (2004).
\newblock \textit{Monte {C}arlo {S}tatistical {M}ethods}.
\newblock 2nd ed. Springer, New York.

\bibitem[{Roberts and Rosenthal(2004)}]{robe:rose:2004}
\textsc{Roberts, G.~O.} and \textsc{Rosenthal, J.~S.} (2004).
\newblock General state space {M}arkov chains and {MCMC} algorithms.
\newblock \textit{Probability Surveys}, \textbf{1} 20--71.

\bibitem[{Rom{\'a}n and Hobert(2012)}]{roma:hobe:2012}
\textsc{Rom{\'a}n, J.~C.} and \textsc{Hobert, J.~P.} (2012).
\newblock Convergence analysis of the {G}ibbs sampler for {B}ayesian general
  linear mixed models with improper priors.
\newblock \textit{The Annals of Statistics}, \textbf{40} 2823--2849.

\bibitem[{Rom{\'a}n and Hobert(2015)}]{roma:hobe:2015}
\textsc{Rom{\'a}n, J.~C.} and \textsc{Hobert, J.~P.} (2015).
\newblock Geometric ergodicity of {G}ibbs samplers for {B}ayesian general
  linear mixed models with proper priors.
\newblock \textit{Linear Algebra and its Applications}, \textbf{473} 54--77.

\bibitem[{Rosenthal(2002)}]{rose:2002}
\textsc{Rosenthal, J.} (2002).
\newblock Quantitative convergence rates of {M}arkov chains: A simple account.
\newblock \textit{Electronic Communications in Probability}, \textbf{7}
  123--128.

\bibitem[{Rosenthal(1995)}]{rose:1995}
\textsc{Rosenthal, J.~S.} (1995).
\newblock Minorization conditions and convergence rates for {M}arkov chain
  {M}onte {C}arlo.
\newblock \textit{Journal of the American Statistical Association}, \textbf{90}
  558--566.

\bibitem[{Roy(2012)}]{roy:2012b}
\textsc{Roy, V.} (2012).
\newblock Convergence rates for {MCMC} algorithms for a robust {B}ayesian
  binary regression model.
\newblock \textit{Electronic Journal of Statistics}, \textbf{6} 2463--2485.

\bibitem[{Roy and Chakraborty(2017)}]{roy:chak:2017}
\textsc{Roy, V.} and \textsc{Chakraborty, S.} (2017).
\newblock Selection of tuning parameters, solution paths and standard errors
  for {B}ayesian lassos.
\newblock \textit{Bayesian Analysis}, \textbf{12} 753--778.

\bibitem[{Roy and Hobert(2007)}]{roy:hobe:2007}
\textsc{Roy, V.} and \textsc{Hobert, J.~P.} (2007).
\newblock Convergence rates and asymptotic standard errors for {M}{C}{M}{C}
  algorithms for {B}ayesian probit regression.
\newblock \textit{Journal of the Royal Statistical Society, {\rm Series B}},
  \textbf{69} 607--623.

\bibitem[{Roy and Hobert(2010)}]{roy:hobe:2010}
\textsc{Roy, V.} and \textsc{Hobert, J.~P.} (2010).
\newblock On {M}onte {C}arlo methods for {B}ayesian regression models with
  heavy-tailed errors.
\newblock \textit{Journal of {M}ultivariate {A}nalysis}, \textbf{101}
  1190--1202.

\bibitem[{Silverman(1986)}]{silv:1986}
\textsc{Silverman, B.~W.} (1986).
\newblock Density estimation for statistics and data analysis.
\newblock \textit{Chapman \& Hall, London}.

\bibitem[{Vats(2017)}]{vats:2017}
\textsc{Vats, D.} (2017).
\newblock Geometric ergodicity of {G}ibbs samplers in {B}ayesian penalized
  regression models.
\newblock \textit{Electronic Journal of Statistics}, \textbf{11} 4033--4064.

\bibitem[{Vats et~al.(2019)Vats, Flegal and Jones}]{vats:fleg:jone:2019}
\textsc{Vats, D.}, \textsc{Flegal, J.~M.} and \textsc{Jones, G.~L.} (2019).
\newblock Multivariate output analysis for {M}arkov chain {M}onte {C}arlo.
\newblock \textit{Biometrka}, \textbf{106} 321--337.

\bibitem[{Vats and Knudson(2018)}]{vats:knud:2018}
\textsc{Vats, D.} and \textsc{Knudson, C.} (2018).
\newblock Revisiting the {G}elman-{R}ubin diagnostic.
\newblock \textit{arXiv preprint arXiv:1812.09384}.

\bibitem[{Wang and Roy(2018{\natexlab{a}})}]{wang:roy:2018a}
\textsc{Wang, X.} and \textsc{Roy, V.} (2018{\natexlab{a}}).
\newblock Analysis of the {P}\'olya-{G}amma block {G}ibbs sampler for
  {B}ayesian logistic linear mixed models.
\newblock \textit{Statistics \& Probability Letters}, \textbf{137} 251--256.

\bibitem[{Wang and Roy(2018{\natexlab{b}})}]{wang:roy:2018}
\textsc{Wang, X.} and \textsc{Roy, V.} (2018{\natexlab{b}}).
\newblock Convergence analysis of the block {G}ibbs sampler for {B}ayesian
  probit linear mixed models with improper priors.
\newblock \textit{Electronic Journal of Statistics}, \textbf{12} 4412--4439.

\bibitem[{Wang and Roy(2018{\natexlab{c}})}]{wang:roy:2018b}
\textsc{Wang, X.} and \textsc{Roy, V.} (2018{\natexlab{c}}).
\newblock Geometric ergodicity of {P}{\'o}lya-{G}amma {G}ibbs sampler for
  {B}ayesian logistic regression with a flat prior.
\newblock \textit{Electronic Journal of Statistics}, \textbf{12} 3295--3311.

\bibitem[{Yu(1994)}]{yu:1994}
\textsc{Yu, B.} (1994).
\newblock Estimating the {L}1 error of kernal estimators: Monitoring
  convergence of {M}arkov samplers.
\newblock \textit{Technical Report, UC Berkeley}.

\end{thebibliography}

\end{document}